\documentclass{article}

\usepackage{arxiv}

\usepackage[utf8]{inputenc} 
\usepackage[T1]{fontenc}    
\usepackage{hyperref}       
\usepackage{url}            
\usepackage{booktabs}       
\usepackage{amsfonts}       
\usepackage{amsmath}
\usepackage{nicefrac}       
\usepackage{microtype}      
\usepackage{optidef}
\usepackage{algpseudocode}
\usepackage{subcaption}
\usepackage{fancybox}
\usepackage{multirow}
\usepackage{graphicx}

\title{Deep Inverse Reinforcement Learning for Structural Evolution of Small Molecules}

\author{
	Brighter Agyemang, \thanks{Contact: brighteragyemang@gmail.com} \\
	School of Computer Science and Engineering\\
	University of Electronic Science and Technology\\
	Chengdu, PRC \\
	   \And
	  Wei-Ping Wu\\
	  School of Computer Science and Engineering\\
	  University of Electronic Science and Technology\\
	  Chengdu, PRC \\
	  \And
	  Daniel Addo\\
	  School of Sofware Engineering\\
	  University of Electronic Science and Technology\\
	  Chengdu, PRC \\
	 \And
	 Michael Y. Kpiebaareh\\
	 School of Computer Science and Engineering\\
	 University of Electronic Science and Technology\\
	 Chengdu, PRC \\
	 \And
	 Ebenezer Nanor\\
	 School of Computer Science and Engineering\\
	 University of Electronic Science and Technology\\
	 Chengdu, PRC \\
	 \And
	 Charles Roland Haruna \\
	 School of Computer Science and Engineering\\
	 University of Electronic Science and Technology\\
	 Chengdu, PRC \\
}

\begin{document}
	\maketitle
	
	\begin{abstract}
	The size and quality of chemical libraries to the drug discovery pipeline are crucial for developing new drugs or repurposing existing drugs. Existing techniques such as combinatorial organic synthesis and High-Throughput Screening usually make the process extraordinarily tough and complicated since the search space of synthetically feasible drugs is exorbitantly huge. While reinforcement learning has been mostly exploited in the literature for generating novel compounds, the requirement of designing a reward function that succinctly represents the learning objective could prove daunting in certain complex domains. Generative Adversarial Network-based methods also mostly discard the discriminator after training and could be hard to train. In this study, we propose a framework for training a compound generator and learning a transferable reward function based on the entropy maximization inverse reinforcement learning paradigm. We show from our experiments that the inverse reinforcement learning route offers a rational alternative for generating chemical compounds in domains where reward function engineering may be less appealing or impossible while data exhibiting the desired objective is readily available.
	\end{abstract}

	\availability{The source code and data of this study are available at https://github.com/bbrighttaer/irelease}
	
	\section{Introduction}
	
	Identifying promising leads is crucial to the early stages of drug discovery. Combinatorial organic synthesis and High-throughput Screening (HTS) are well-known methods used to generate new compounds in the domain (drug and compound are used interchangeably in this study). This generation process is typically followed by expert analysis, which focuses on desired properties such as solubility, activity, pharmacokinetic profile, toxicity, and synthetic accessibility, to ascertain the desirability of a generated compound. Compound generation and modification methods are useful for enriching chemical databases and scaffold hopping~\cite{Olivecrona2017}. A review of the structural and analog entity evolution patent landscape estimates that the pharmaceutical industry constitutes $70\%$ of the domain~\cite{Ivanenkov2017}.
	
	Indeed, the compound generation task is noted to be hard and complicated~\cite{Segler2018} considering that there exist $10^{30}-10^{60}$ synthetically feasible drug-like compounds~\cite{Polishchuk2013}. With about $96\%$~\cite{Hingorani2019} of drug development projects failing due to unforeseen reasons, it is significant to ensure diversity in compounds that are desirable to avoid a fatal collapse of the drug discovery process. As a result of these challenges in the domain, there is a need for improved de novo compound generation methods.
	
	In recent times, the proliferation of data, advances in computer hardware, novel algorithms for studying complex problems, and other related factors have contributed significantly to the steady growth of data-driven methods such as Deep Learning (DL). DL-based approaches have been applied to several domains, such as Natural Language Processing (NLP)~\cite{Bahdanau2014, Radford2018}, Computer Vision~\cite{Xu2015}, ProteoChemometric Modeling~\cite{Ragoza2017}, compound and target representation learning~\cite{Agyemang2020, Tsubaki2019}, and reaction analysis~\cite{Segler2017}. Consequently, there has been a growing interest in the literature to use data-driven methods to study the compound generation problem.
	
	Deep Reinforcement Learning (DRL), Generative Adversarial Networks (GAN), and Transfer Learning (TL) are some of the approaches that have been used to generate compounds represented using the Simplified Molecular Input Line Entry System (SMILES)~\cite{Weininger1988}. The DRL-based methods model the compound generation task as a sequential decision-making process and use Reinforcement Learning (RL) algorithms to design generators (agents) that estimate the statistical relationship between actions and outcomes. This statistical knowledge is then leveraged to maximize the outcome, thereby biasing the generator according to the desired chemical space. Motivated by the work in~\cite{Yu2016}, the GAN-based methods also model the compound generation task as a sequential decision-making problem but with a discriminator parameterizing the reward function. TL methods train a generator on a large dataset to increase the proportion of valid SMILES strings sampled from the generator before performing a fine-tuning training to bias the generator to the target chemical space.
	
	Regarding DRL-based methods, ~\cite{Olivecrona2017} proposed a Recurrent Neural Network (RNN)-based approach to train generative models for producing analogs of a query compound and compounds that satisfy certain chemical properties, such as activity to the Dopamine Receptor D2 (DRD2) target. The generator takes as input one-hot encoded representations of the canonical SMILES of a compound, and for each experiment, a corresponding reward function is specified. Also,~\cite{Popova2018} proposed a stack-augmented RNN generative model using the REINFORCE~\cite{Williams1992} algorithm where, unlike~\cite{Olivecrona2017}, the canonical SMILES encoding of a compound is learned using backpropagation. The reward functions in~\cite{Popova2018} are parameterized by a prediction model that is trained separately from the generator. In both the studies of~\cite{Olivecrona2017} and~\cite{Popova2018}, the generator was pretrained on a large SMILES dataset using a supervised learning approach before applying RL to bias the generator. Similarly,~\cite{Shi2019} proposed a SMILES- and Graph-based compound generation model that adopts the supervised pretraining and subsequent RL biasing approach. Unlike~\cite{Olivecrona2017} and~\cite{Popova2018},~\cite{Shi2019} assign intermediate rewards to valid incomplete SMILES strings. We posit that since an incomplete SMILES string could, in some cases, have meaning (e.g., moieties), assigning intermediate rewards could facilitate learning. While the DRL-based generative models can generate biased compounds, an accurate specification of the reward function is challenging and time-consuming in complex domains. In most interesting de novo compound generation scenarios, compounds meeting multiple objectives may be required, and specifying such multi-objective reward functions leads to the generator (agent) exploiting the straightforward objective and generating compounds with low variety. 
	
	In another vein,~\cite{Segler2018} trained an RNN model on a large dataset using supervised learning and then performed TL to the domain of interest to generate focused compounds. Since the supervised approach used to train the generator is different from the autoregressive sampling technique adopted at test time, such methods are not well suited for multi-step SMILES sampling~\cite{Schmidt2016}. This discrepancy is referred to as exposure bias. Additionally, methods such as~\cite{Segler2018} that maximize the likelihood of the underlying data are susceptible to learning distributions that place masses in low-density areas of the multivariate distribution giving rise to the underlying data.
	
	On the other hand,~\cite{Sanchez-Lengeling2017} based on the work of~\cite{Guimaraes2017} to propose a GAN-based generator that produces compounds matching some desirable metrics. The authors adopted an alternating approach to enable multi-objective RL optimization. As pointed out by~\cite{Benhenda2017}, the challenges with training GAN cause a high rate of invalid SMILES strings, low diversity, and reward function exploitation. In~\cite{Putin2018}, a memory-based generator replaced the generator proposed by~\cite{Sanchez-Lengeling2017} in order to mitigate the problems in~\cite{Sanchez-Lengeling2017}. In these GAN-based methods, the authors adopted a Monte-Carlo Tree Search (MCTS) method to assign intermediate rewards. However, GAN training could be unstable and the generator could get worse as a result of early saturation~\cite{Arjovsky2019}. Additionally, the discriminator in the GAN-based models is typically discarded after training. 
	
	In this paper, we propose a novel framework for training a compound generator and learning a reward function from data using a sample-based Deep Inverse Reinforcement Learning (DIRL) objective~\cite{Finn2016a}. We observe that while it may be daunting or impossible to accurately specify the reward function of some complex drug discovery campaigns to leverage DRL algorithms, sample of compounds that satisfy the desired behavior may be readily available or collated. Therefore, our proposed method offers a solution to developing in-silico generators and recovering reward function from compound samples. As pointed out by~\cite{Finn2016b}, the DIRL objective could lead to stability in training and producing effective generators. Also, unlike the typical GAN case where the discriminator is discarded, the sample-based DIRL objective is capable of training a generator and a reward function. Since the learned reward function succinctly represents the agent's objective, it could be transferred to related domains (with a possible fine-tuning step). Moreover, since the Binary Cross Entropy (BCE) loss usually applied to train a discriminator does not apply in the case of the sampled-based DIRL objective, this eliminates saturation problems in training the generator. The DIRL approach also mitigates the challenge of imbalance between different RL objectives.
	
	The outline of our study is as follows: Section~\ref{sec:prelims} presents the RL and IRL background of this study, Section~\ref{sec:methods} discusses the research problem of this study and our proposed approach, we discuss the results of our experiments in Section~\ref{sec:discussion}, and draw the conclusions of this study in Section~\ref{sec:conclusion}.

	
	\section{Preliminaries}\label{sec:prelims}
	In this section, we review the concepts of Reinforcement Learning (RL) and Inverse Reinforcement Learning (IRL) related to this study.
	
	\subsection{Reinforcement Learning}\label{subsec:rl}
	The aim of Artificial Intelligence (AI) is to develop autonomous agents that can sense their environment and act intelligently. Since learning through interaction is vital for developing intelligent systems, the paradigm of RL has been adopted to study several AI research problems. In RL, an agent receives an observation from its environment, reasons about the received observation to decide on an action to execute in the environment, and receives a signal from the environment as to the usefulness of the action executed in the environment. This process continues in a trial-and-error manner over a finite or infinite time horizon for the agent to estimate the statistical relationship between the observations, actions, and their results. This statistical relationship is then leveraged by the agent to optimize the expected signal from the environment.
	
	Formally, an RL agent receives a state $s_t$ from the environment and takes an action $a_t$ in the environment leading to a scalar reward $r_{t+1}$ in each time step $t$. The agent's behavior is defined by a policy $\pi(a_t|s_t)$, and each action performed by the agent transitions the agent and the environment to a next state $s_{t+1}$ or a terminal state $s_T$.
	The RL problem is modeled as a Markov Decision Process (MDP) consisting of:
	\begin{itemize}
		\item A set of states $\mathcal{S}$, with a distribution over starting states $p(s_0)$.
		\item A set of actions $\mathcal{A}$.
		\item State transition dynamics function $\mathcal{T}(s_{t+1}|s_t, a_t)$ that maps a state-action pair at a time step $t$ to a distribution of states at time step $t+1$.
		\item A reward function $\mathcal{R}(s_t,a_t)$ that assigns a reward to the agent after taking action $a_t$ in state $s_t$.
		\item A discount factor $\gamma\in [0,1]$ that is used to specify a preference for immediate and long-term rewards. Also, $\gamma < 1$ ensures that a limit is placed on the time steps considered in the infinite horizon case.
	\end{itemize}
	The policy $\pi$ maps a state to a probability distribution over the action space: $\pi: S \rightarrow p(\mathcal{A}=a|\mathcal{S})$. In an episodic MDP of length $T$, the sequence of states, actions, and rewards is referred to as a trajectory or rollout of the policy $\pi$. The return or total reward for a trajectory could then be represented as $R=\sum_{t=0}^{T-1}\gamma^tr_{t+1}$. The goal of an RL agent is to learn a policy $\pi^*$ that maximizes its expected total reward from all states:
	\begin{equation}
		\pi^* = \underset{\pi}{argmax}\mathbb{E}\left[R|\pi\right]
	\end{equation}
	Existing RL methods for solving such an MDP could be classified into value function-based and policy search algorithms. Value function methods estimate the expected return of being in a given state. The state-value function $V^\pi(s)$ estimates the expected return of starting in state $s$ and following $\pi$ afterward:
	\begin{equation}
		V^\pi(s) = \mathbb{E}\left[R|s,a,\pi\right]
	\end{equation}
	Given the optimal values of all states, $V^*(s)=\underset{\pi}{max}V^\pi(s), \forall s\in S$, the optimal policy $\pi^*$ could be determined as:
	\begin{equation}
		\pi^*=\underset{\pi}{argmax}V^*(s)
	\end{equation}
	A closely related concept to the state-value function is the state-action value function $Q^\pi(s,a)$. With the state-action value function, the initial action $a$ is given and the policy $\pi$ is effective from the next state onward:
	\begin{equation}
		Q^\pi(s,a) = \mathbb{E}\left[R|s,a,\pi\right].
		\label{eq:q_value}
	\end{equation}
	The agent can determine the optimal policy, given $Q^\pi(s,a)$, by acting greedily in every state, $\pi^*=\underset{\pi}{argmax} Q^\pi(s,a)$. Also, $V^\pi(s)=\underset{a}{max}Q^\pi(s,a)$.
	
	On the other hand, policy search/gradient algorithms directly learn the policy $\pi$ of the agent instead of indirectly estimating it from value-based methods. Specifically, policy-based methods optimize the policy $\pi(a|s,\theta)$, where $\theta$ is the set of parameters of the model approximating the true policy $\pi^*$. We review the Policy Gradient (PG) methods used in this study below.
	
	\subsubsection{Policy Gradient Optimization}
	The REINFORCE algorithm~\cite{Williams1992} is a policy gradient method that estimates the gradient $g := \nabla_\theta\mathbb{E}\left[\sum_{t=0}^{T-1}r_{t+1}\right]$ which has the form
	\begin{equation}
		g = \mathbb{E}\left[\sum_{t=0}^{T-1}\Psi_t\nabla_\theta log\pi_\theta(a_t|s_t)\right],
		\label{eq:policy_gradient}
	\end{equation}
	and updates $\theta$ in the direction of the estimated gradient. $\Psi_t$ could be any of the functions defined by~\cite{Schulmanetal_ICLR2016}. For instance,~\cite{Popova2018} defined $\Psi_t=\gamma^t r(s_T)$, where $r(s_T)$ denotes the reward at the terminal state. In another vein, ~\cite{Putin2018} used the reward of a fully generated sequence. Due to high variance in using the total reward of the trajectory $\sum_{t=0}^{T-1}\gamma^t{r_{t+1}}$, a baseline version could be adopted to reduce the variance, $\sum_{t'=t}^{T-1}\gamma^{t'-t}r_{t'+1}-b(s_t)$. When a value function approximator is used to estimate the baseline $b(s_t)$ the resulting method is referred to as an actor-critic method. A related concept to determining $\Psi_t$ is realizing that it is less challenging to identify that an action has a better outcome than the default behavior compared to learning the value of the action~\cite{Arulkumaran2017}. This concept gives rise to the advantage function $A^\pi(s,a) = Q^\pi(s,a) - V^\pi(s)$ where $V(s)$, serving as a baseline, is approximated by a function such as a neural network.
	
	\subsubsection{Proximal Policy Optimization}
	Schulman et al.~\cite{Schulmanppo} proposed a robust and data efficient policy gradient algorithm as an alternative to prior RL training methods such as the REINFORCE, Q-learning~\cite{Mnih2015}, and the relatively complicated Trust Region Policy Optimization (TRPO)~\cite{Schulman2015}. The proposed method, named Proximal Policy Optimization (PPO), shares some similarities with the TRPO in using the ratio between the new and old policies scaled by the advantages of actions to estimate the policy gradient instead of the logarithm of action probabilities (as seen in Equation~\ref{eq:policy_gradient}). While TRPO uses a constrained optimization objective that requires the conjugate gradient algorithm to avoid large policy updates, PPO uses a clipped objective that forms a pessimistic estimate of the policy's performance to avoid destructive policy updates. These attributes of the PPO objective enables performing multiple epochs of policy update with the data sampled from the environment. In REINFORCE, such multiple epochs of optimization often leads to destructive policy updates. Thus, PPO is also more sample efficient than the REINFORCE algorithm.
	
	More formally, the PPO objective is expressed as:
	\begin{equation}
		L^{PPO}\left(\theta\right) = \mathbb{E}\left[min\left(r_t(\theta)A_t, clip(r_t(\theta), 1-\epsilon, 1+\epsilon)A_t\right)\right]
	\end{equation}
	where
	\begin{equation}
		r_t(\theta)=\frac{\pi_\theta(a_t|s_t)}{\pi_{\theta_{old}}(a_t|s_t)},
	\end{equation}
	and
	\begin{equation}
		A_t = \delta_t + (\gamma\lambda)\delta_{t+1}+...+(\gamma\lambda)^{T-t+1}\delta_{T-1}
	\end{equation}
	where $\delta_t=r_t+\gamma V(s_{t+1})-V(s_t)$ with hyperparameters $\gamma$ and $\lambda$. The $clip(\cdot)$ function creates an alternate policy update to the expectation of $r_t(\theta)A_t$ where the action probability ratios between the old and updated policies are maintained in the range of $[1-\epsilon, 1+\epsilon]$. Taking the minimum of the two possible policy updates ensures a lower bound update is preferred. This is useful since a large update to the parameters of a highly non-linear policy function, such as a neural network, often results in a worse policy. In this paper, we set $\epsilon=0.2$.
	
	\subsection{Inverse Reinforcement Learning}\label{subsec:irl}
	IRL is the problem of learning the reward function of an observed agent, given its policy or behavior, thereby avoiding the manual specification of a reward function~\cite{Arora2018}. The IRL class of solutions assumes the following MDP$\setminus R_E$:
	\begin{itemize}
		\item A set of states $\mathcal{S}$, with a distribution over starting states $p(s_0)$.
		\item A set of actions $\mathcal{A}$.
		\item State transition dynamics function $\mathcal{T}(s_{t+1}|s_t, a_t)$ that maps a state-action pair at a time step $t$ to a distribution of states at time step $t+1$.
		\item A set of demonstrated trajectories \\ $\mathcal{D}=\left\{\left<(s_0^i,a_0^i),...,(s_{T-1}^i,a_{T-1}^i)\right>_{i=1}^N\right\}$ from the observed agent or expert. 
		\item A discount factor $\gamma\in [0,1]$ may be used to discount future rewards.
	\end{itemize}
	The goal of IRL then is to learn the reward function $R_E$ that best explains the expert demonstrations. It is assumed that the demonstrations are perfectly observed and that the expert follows an optimal policy.
	
	While learning reward functions from data is appealing, the IRL problem is ill-posed since there are many reward functions under which the observed expert behavior is optimal~\cite{Ng2000,Arora2018}. An instance is a reward function that assigns 0 (or any constant value) for all selected actions; in such a case, any policy is optimal. Other main challenges are accurate inference, generalizability, the correctness of prior knowledge, and computational cost with an increase in problem complexity.
	
	To this end, several IRL proposals exist in the literature to mitigate the IRL challenges mentioned above~\cite{Arora2018}. In recent times, the entropy optimization class of IRL methods has been widely used by researchers due to the maximum entropy's goal to obtain an unbiased distribution of potential reward~\cite{Ziebart2008, Wulfmeier2016, Ho2016, Finn2016a}. The intuition is that the solution that maximizes entropy violates the optimization constraints least and hence, least wrong. In the Maximum Entropy (MaxEnt) formulation, the probability of the expert's trajectory is proportional to the exponential of the total reward~\cite{Ziebart2008},
	\begin{equation}
		p(\tau) \propto exp(R(\tau)), \forall \tau\in\mathcal{D},
	\end{equation}
	where $R(\tau)=\sum_{(s,a)\in\tau}r(s,a)$ and $r(s,a)$ gives the reward for taking action $a$ in state $s$.

		\section{Methods}\label{sec:methods}
		\subsection{Problem Statement}\label{sec:problem}
		We consider the problem of training a model $G_\theta(a_t|s_{0:t})$ to generate a set of compounds $\mathbf{C}=\left\{c_1, c_2, ..., c_M|M\in\mathbb{N}\right\}$, each encoded as a valid SMILES string, such that $\mathbf{C}$ is generally biased to satisfy a predefined criteria that could be evaluated by a function $E$. Considering the sequential nature of a SMILES string, we study this problem in the framework of IRL following the MDP described in section~\ref{subsec:irl}; instead of the RL approach mostly adopted in the literature on compound generation, a parameterized reward function $R_\psi$ has to be learned from $\mathcal{D}$, which is a set of SMILES that satisfy the criteria evaluated by $E$, to train $G_\theta$ following the MDP in section~\ref{subsec:rl}. Note that in the case of SMILES generation, $\mathcal{T}(s_{t+1}|s_t, a_t)=1$.
		
		In this context, the action space $\mathcal{A}$ is defined as the set of unique tokens that are combined to form a canonical SMILES string and the state space $\mathcal{S}$ is the set of all possible combinations of these tokens to encode a valid SMILES string, each of length $l\in [1, T], T\in\mathbb{N}$. We set $s_0$ to a fixed state that denotes the beginning of a SMILES string, and $s_T$ is the terminal state. 
		
		Furthermore, we compute $r(s_{T-1},a_{T-1})=R_\psi(Y_{1:T})$, where $Y_{1:T}$ denotes a full generated sequence. Intermediate state rewards are obtained by performing an $N$-time Monte Carlo Tree Search (MCTS) with $G_\theta$ as the rollout policy to obtain the corresponding rewards. Specifically, given the $N$-time MCTS set of rollouts,
		\begin{equation}
			MC^{G_\theta}(Y_{1:t}, N) = \left\{Y^1_{1:T}, ..., Y^N_{1:T}\right\},
		\end{equation}
		the reward for a state is then calculated as
		\[ r(s_t, a_t) =
		\begin{cases}
			\frac{1}{N}\sum_{n=1}^N R_\psi(Y_{1:T}^n), Y_{1:T}^n \in MC^{G_\theta}(Y_{1:t}, N)\quad t<T-1,\\
			R_\psi(Y_{1:T}), \quad t=T-1.
		\end{cases}
		\]
		
		\subsection{Proposed Approach}
\begin{figure*}[]
			\centering
			\subcaptionbox{\label{fig:framework}}{
				\includegraphics[scale=0.8]{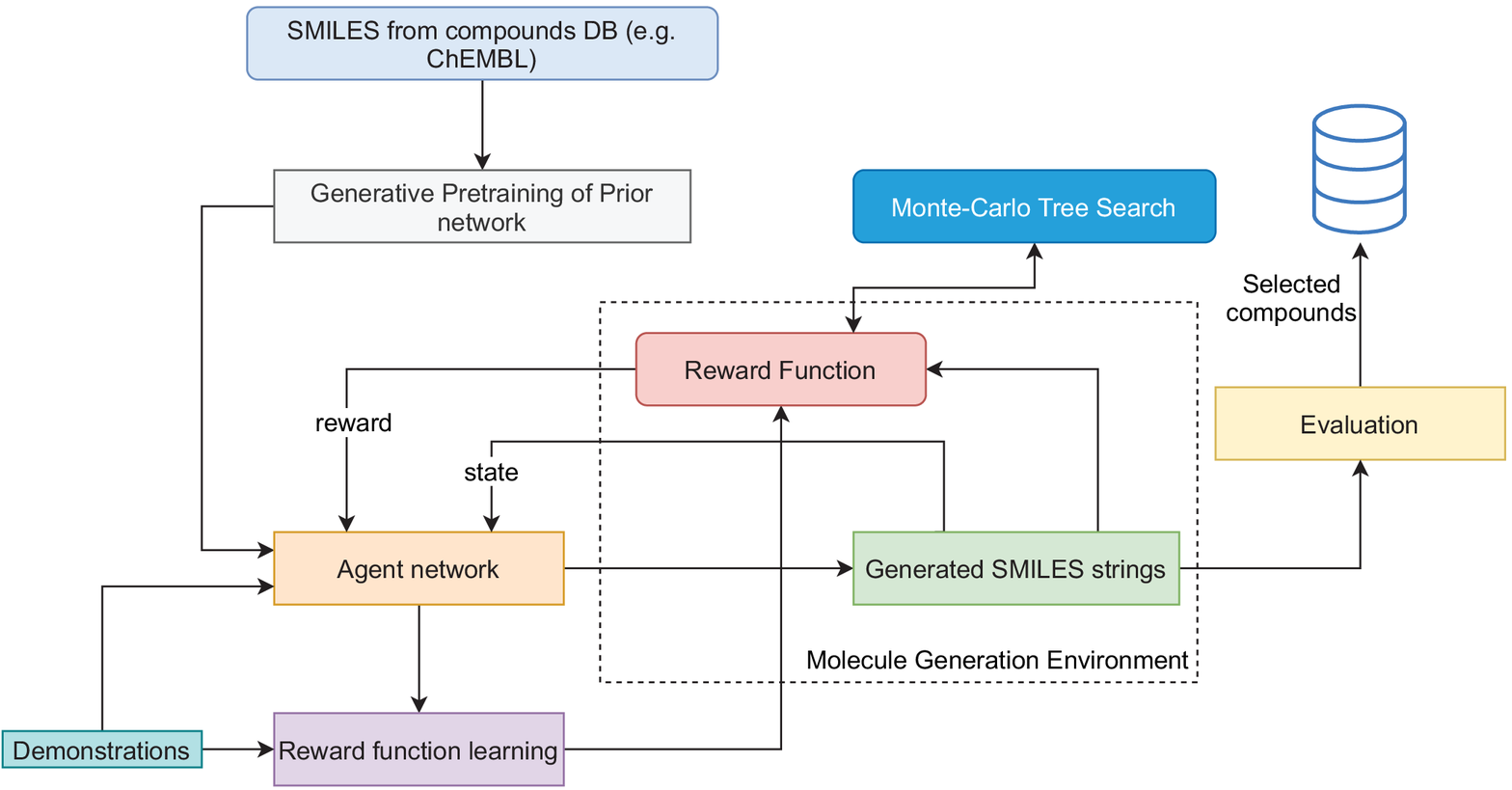}
			}\hspace{1em}%
			\\
			\subcaptionbox{\label{fig:modelsarch}}{
				\includegraphics[scale=0.6]{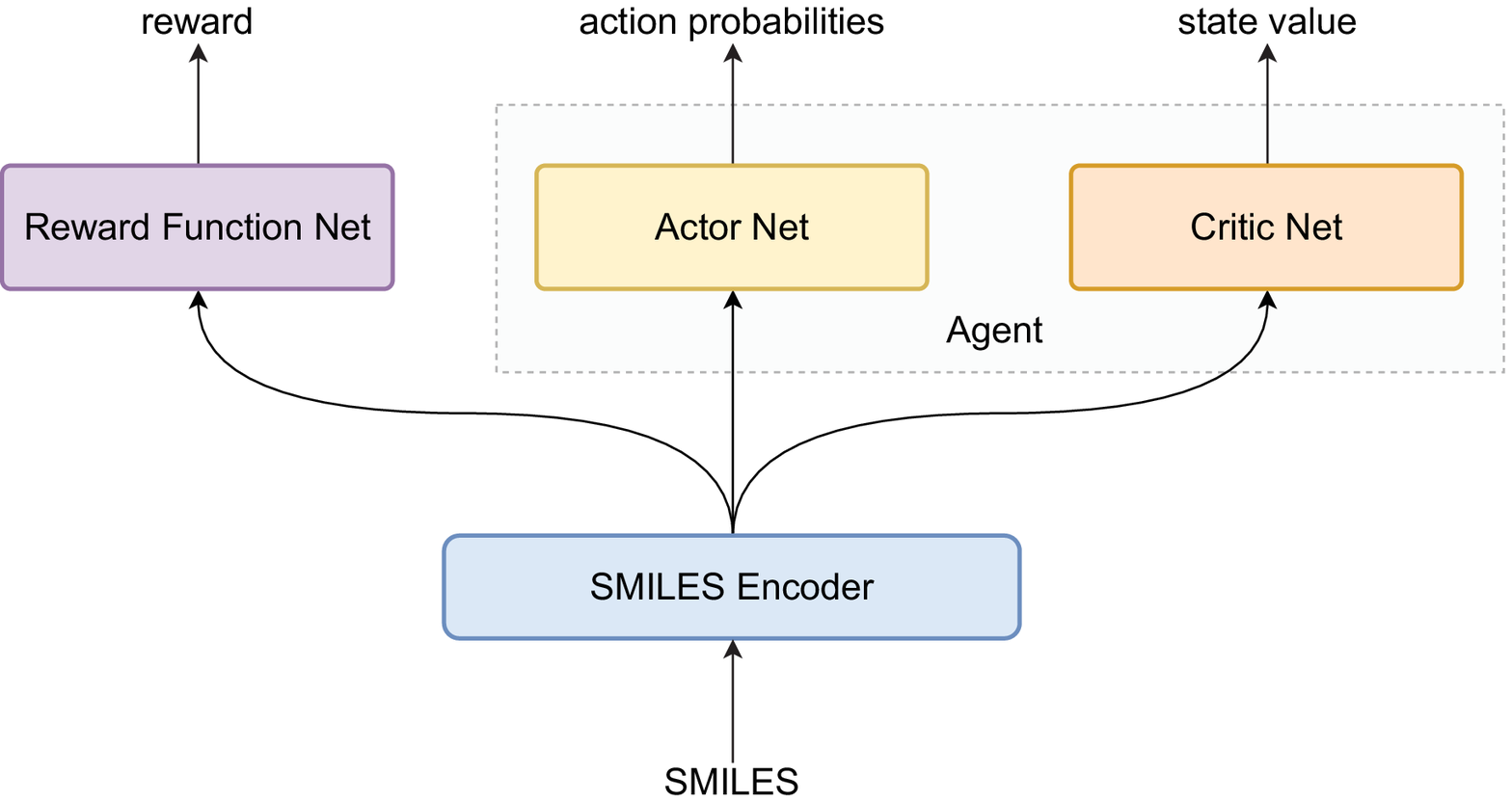}
			}\hspace{1em}
			\label{fig:framework_and_models_arch}
			\caption{(a) Illustration of the proposed framework for training a small molecule generator and learning a reward function using IRL. The workflow begins with pretraining the generator model using large-scale compound sequence dataset, such as ChEMBL. The pretrained model is used to initialize the agent network in an IRL training scheme where a reward function is learned and the agent is biased to generate desirable compounds. The generated SMILES could be examined by an evaluation function where compounds satisfying a specified threshold could be persisted for further analysis. (b) The general structure of the models used in the study. The agent is represented as an actor-critic architecture that is trained using the PPO~\cite{Schulmanppo} algorithm. The actor, critic, and reward function are all RNNs that share the same SMILES encoder. The actor net becomes the SMILES generator at test time.}
		\end{figure*}

		The workflow we propose in this study for the structural evolution of compounds is presented in Figure~\ref{fig:framework}. The framework is described in what follows.
		
		\subsection{Datasets}\label{subsec:datasets}
		The first stage of the workflow is creating a SMILES dataset, similar to most existing GAN and vanilla-RL methods. Possible repositories for facilitating this dataset creation are DrugBank~\cite{Knox2011}, KEGG~\cite{Kanehisa2012}, STITCH~\cite{Szklarczyk2016}, and ChEMBL~\cite{Bento2014}. This dataset is used at the next stage for pretraining the generator. For this purpose, we used the curated ChEMBL dataset of~\cite{Popova2018}, which consists of approximately $1.5$ million drug-like compounds.
		
		Also, a set of SMILES satisfying the constraints or criteria of interest are collated from an appropriate source as the demonstrations $\mathcal{D}$ of the IRL phase (for instance, see section~\ref{sec:exp_setup}). To evaluate SMILES during the training of the generator, we assume the availability of an evaluation function $ E $ that can evaluate the extent to which a given compound satisfies the optimization constraints. Here, $E$ could be an ML model, a robot that conducts chemical synthesis, or a molecular docking program.
		
		In this study, we avoided composing the demonstrations dataset from the data used to train the evaluation function. Since, in practice, the data, rule set, or method used for developing any of the evaluation functions mentioned above could differ from the approach for constructing the demonstrations dataset, this independence provides a more realistic context to assess our proposed method.
		
		\subsection{Generative Pretraining of Prior Model}\label{sec:prior_model_pretraining}
		This stage of the framework entails using the large-scale dataset of the previous step to pretrain a model that would be a prior for initializing the agent network/policy. This pretraining step aims to enable the model to learn the SMILES syntax to attain a high rate of valid generated SMILES strings. 
		
		Since the SMILES string is a sequence of characters, we represent the prior model as a Recurrent Neural Network (RNN) with Gated Recurrent Units (GRUs). The architecture of the generator at this stage of the workflow is depicted in Figure~\ref{fig:agent_net_train}. The generator takes as input the output of an encoder that learns the embedding of SMILES characters, as shown in Figure~\ref{fig:modelsarch}. For each given SMILES string, the generator is trained to predict the next token in the sequence using the cross-entropy loss function.
		
		According to the findings of~\cite{Joulin2015}, regular RNNs cannot effectively generate sequences of a context-free language due to the lack of memory. To this end,~\cite{Joulin2015} proposed the use of a Stack-RNN architecture, which equips the standard RNN cell with a stack unit to mitigate this problem. Since the SMILES encoding of a compound is a context-free language, we follow~\cite{Popova2018} to use the Stack-RNN for generating SMILES as it ensures that tasks such as noting the start and end parts of aromatic moieties, and parentheses for branches in the chemical structure are well addressed. We note that while~\cite{Popova2018} used a single layer Stack-RNN, we adopt a multi-layer Stack-RNN structure in our study. We reckon that the multi-layer Stack-RNN could facilitate learning better representations at multiple levels of abstraction akin to multi-layer RNNs.
		
		The stack unit enables the model to persist information across different time steps. Three differentiable operations could be performed on the stack at each time step: POP, PUSH, and NO-OP. The POP operation deletes the entry at the top of the stack, the PUSH operation updates the top of the stack with a new record, and the NO-OP operation leaves the stack unchanged for the next time step. In~\cite{Joulin2015}, the entry at the top of stack $s$ at time step $t$ is indexed as $s_t[0]$ and is updated as
		\begin{equation}
			s_t[0] = a_t[PUSH]\sigma(Dh_t) + a_t[POP]s_{t-1}[1] + a_t[NO-OP]s_{t-1}[0],
		\end{equation}
		where $h_t\in\mathbb{R}^{m\times 1}$ is the RNN hidden vector of a sequence at time step $t$, $D\in\mathbb{R}^{w\times m}$, $w$ is the dimension of the stack, and $a_t\in\mathbb{R}^3$ is a vector whose elements map to the stack control operators PUSH, POP, and NO-OP, respectively. Here, $a_t$ is computed as 
		\begin{equation}
			a_t = softmax(Ah_t).
		\end{equation}
		where $A\in\mathbb{R}^{3\times m}$. Intuitively, if $a_t[PUSH]=1$ then $s_t[0]$ is updated with $Dh_t$; if $a_t[POP]=1$ then the top of the stack is updated with the second entry of the stack and $a_t[NO-OP]=1$ leaves the stack unchanged.
		
		Regarding the remaining entries of the stack, thus indexes $i>0$, the update rule is
		\begin{equation}
			s_t[i]=a_t[PUSH]s_{t-1}[i-1] + a_t[POP]s_{t-1}[i+1] + a_t[NO-OP]s_{t-1}[i].
		\end{equation}
		Subsequently, the RNN cell's hidden vector $h_t$ is updated as
		\begin{equation}
			h_t=\sigma(Ux_t + Rh_{t-1} + Ps_{t-1}[0]),
		\end{equation}
		where $\sigma(x)=1/(1+exp(-x))$, $U\in\mathbb{R}^{m\times d}$, $R \in\mathbb{R}^{m\times m}$, $P\in\mathbb{R}^{m\times w}$, $s_{t-1}[0]\in\mathbb{R}^{w\times 1}$, $d$ is the dimension of encoded representation of token $x_t$, and $m$ is dimension of the hidden state.
		
		\begin{figure*}
			\centering
			\includegraphics[width=0.8\linewidth]{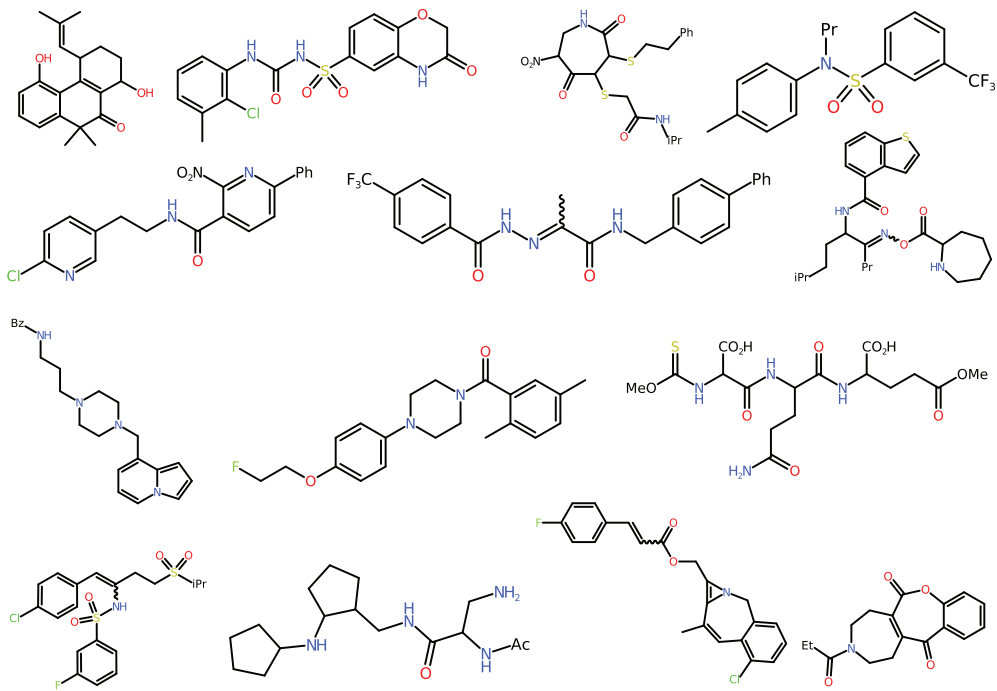}
			\caption{Sample compounds generated using the pretrained model.}
			\label{fig:cdkunbiasedsmiles}
		\end{figure*}

		\subsection{Inverse Reinforcement Learning Phase}
		As stated earlier, we frame the SMILES string generation task as an MDP problem where the reward function is unknown. Consequently, this MDP does not enable the use of RL algorithms to approximate the solution since RL methods require a reward function. Therefore we learn the reward function using demonstrations that exhibit the desired behavior.
		
		Specifically, given the set of trajectories $D=\left\{\tau_1,\tau_2,...\tau_N\right\}$ observed from an expert that acts stochastically under an optimal policy, the density function is specified as,
		\begin{equation}
			p(\tau) = \frac{1}{Z}exp(R\psi(\tau))
			\label{eqn:p_tau}
		\end{equation}
		where $\tau=\left\{\left<s_0,a_0\right>, ..., \left<s_{T-1},a_{T-1}\right>\right\}$ is a trajectory of a generated SMILES string $Y_{1:T}$ and $R\psi$ is an unknown reward function parameterized by $\psi$. Thus, the expert generates SMILES strings that satisfy the desired criteria with a probability that increases exponentially with the return. Here, the main challenge for this energy-based model is computing the partition function,
		\begin{equation}
			Z = \int exp(R_\psi(\tau)).
		\end{equation}
		Ziebart et al.~\cite{Ziebart2008} computed $Z$ using a Dynamic Programming (DP) approach to estimate the partition function. However, the DP method is not scalable to large-state space domains, such as the space of synthetically accessible compounds mentioned earlier. 
		

		\begin{figure*}[]
			\centering
			\subcaptionbox{\label{fig:reward_critic_nets}}{
				\includegraphics[width=.5\linewidth]{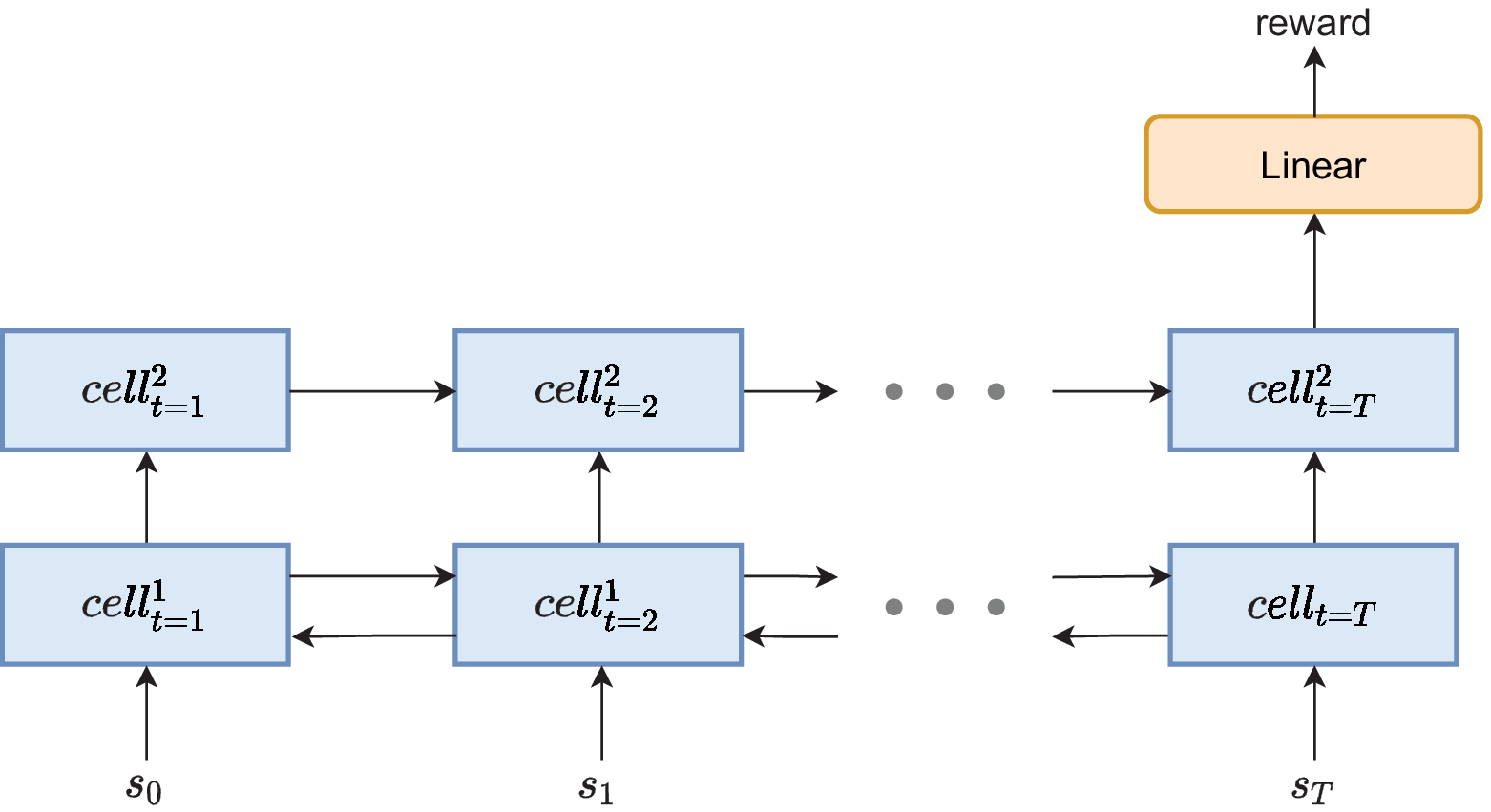}
				\hspace{2em}
				\includegraphics[width=.5\linewidth]{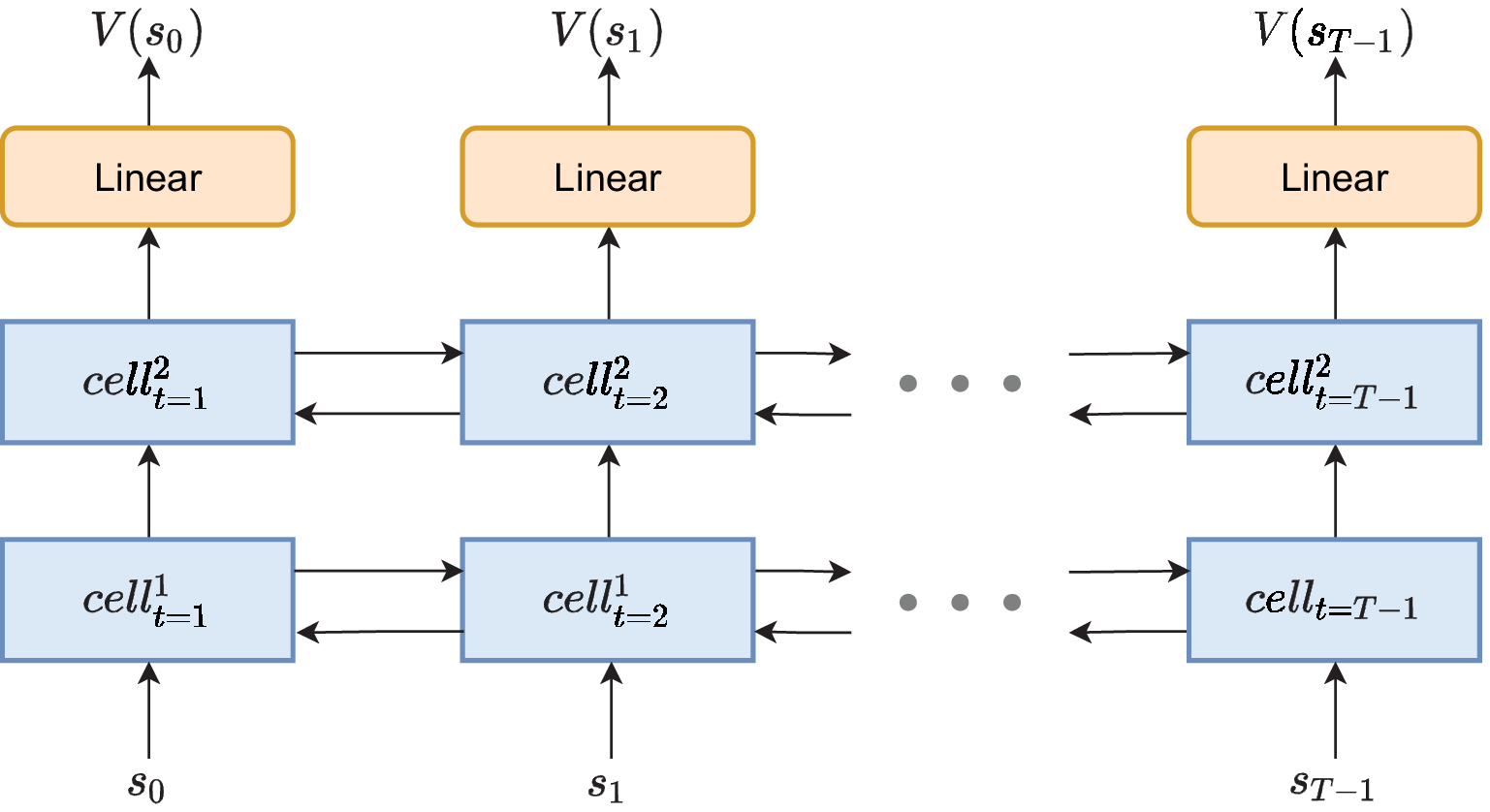}
			}\hspace{1em}%
			\\
			\subcaptionbox{\label{fig:agent_net_train}}{\includegraphics[width=\linewidth]{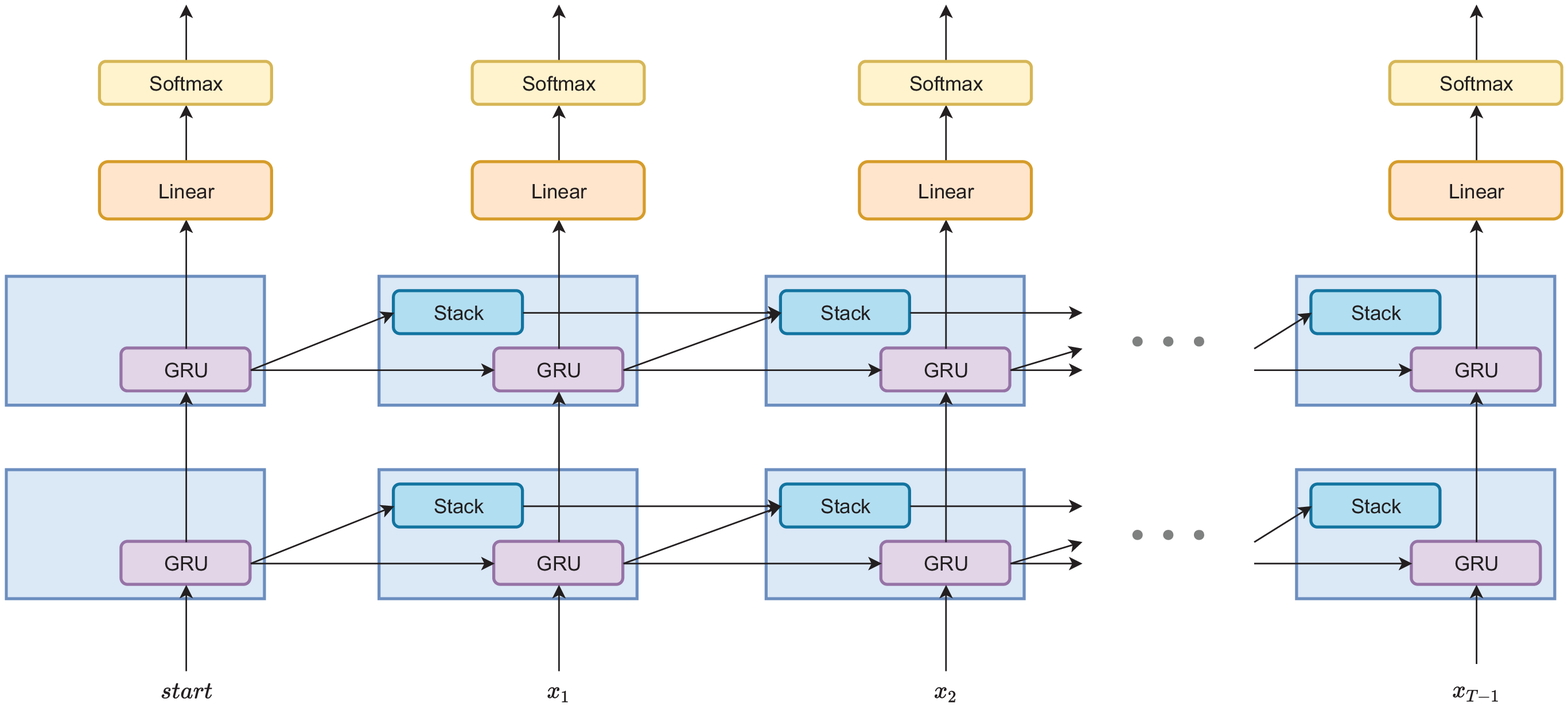}}\hspace{1em}%
			\\
			\subcaptionbox{\label{fig:agent_net_eval}}{\includegraphics[width=\linewidth]{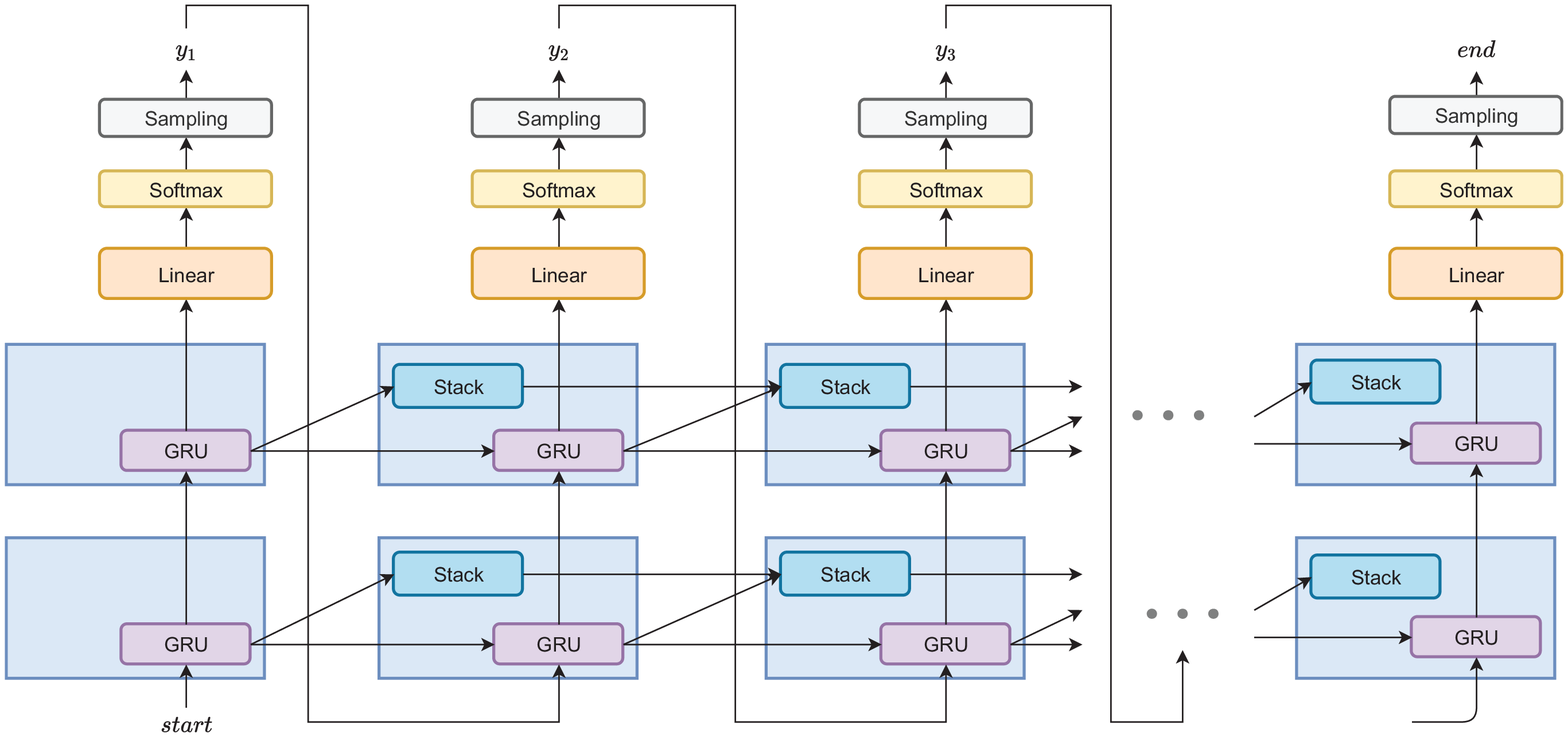}}\hspace{1em}%
			\label{fig:detai_modelsarch}
			\caption{The architectures of the RNN models used in this study. (a) The structure of the reward net (left) and the critic net (right). (b) The structure of the generator net during pretraining to learn the SMILES syntax. The model is trained using the cross-entropy loss (c) The structure of the generator net during the RL phase; the model autoregressively generates a SMILES string given a fixed start state. The sampling process terminates when an $end$ token is sampled, or a length limit is reached.}
		\end{figure*}
		
		In~\cite{Finn2016a}, the authors proposed a sample-based entropy maximization IRL algorithm. The authors use a sample or background distribution to approximate the partition function and continuously adapt this sample distribution to provide a better estimation of Equation~\ref{eqn:p_tau}. Therefore, if the sample distribution represents the agent's policy and this distribution is trained using an RL algorithm, given an approximation of $R_\psi$, then the RL algorithm guides the sample distribution to a space that provides a better estimation of $Z$. We refer to the method discussed by~\cite{Finn2016a} as Guided Reward Learning (GRL) in this study. We adopt this approach to bias a pretrained generator to the desired compound space since the GRL method produces a policy and a learned reward function that could be transferred to other related domains.
		
		Formally, taking the log-likelihood of Equation~\ref{eqn:p_tau} provides the maximization objective,
		\begin{align}
			\mathcal{L}_{GRL}(\psi) &= \frac{1}{N}\sum_{\tau_i\in \mathcal{D}}R_\psi(\tau_i) - log Z \\
			&\approx \frac{1}{N}\sum_{\tau_i\in\mathcal{D}}R_\psi(\tau_i) - log \frac{1}{M}\sum_{\tau_j\in\hat{\mathcal{D}}}\frac{exp(R_\psi(\tau_j))}{q(\tau_j)}
			\label{eq:grl_objective}
		\end{align}
		where $\hat{\mathcal{D}}$ is the set of $M$ trajectories sampled using the background distribution $q$. As depicted in Figure~\ref{fig:reward_critic_nets}-left, $R_\psi$ is represented as a 2-layer GRU RNN that takes as input the output of a SMILES encoder (see Figure~\ref{fig:modelsarch}) and predicts a scalar reward for a generated valid or invalid SMILES string.
		
		RL is used to train $q$ to improve the background distribution used in estimating the partition function. The model architecture of $q$ during the RL training and SMILES generation phases is shown in Figure~\ref{fig:agent_net_eval}.
		
		Since we represent $q$ as an RNN in this study, training the sample distribution with high-variance RL algorithms such as the REINFORCE~\cite{Williams1992} objective could override the efforts of the pretraining due to the non-linearity of the model. Therefore, we train the sample distribution using the Proximal Policy Optimization (PPO) algorithm~\cite{Schulmanppo}. The PPO objective ensures a gradual change in the parameters of the agent policy (sample distribution).
		The PPO algorithm learns a value function to estimate $A^\pi(s, a)$ making it an actor-critic method. The architecture of the critic in this study, modeled as an RNN, is illustrated in~\ref{fig:reward_critic_nets}-right. Like the actor model, the critic takes the SMILES encoder's outputs as input to predict $V(s_t)$ at each time step $t$.
		
		Lastly, the evaluation function $E$ of our proposed framework (see Figure~\ref{fig:framework}) is used, at designated periods of training, to evaluate a set of generated SMILES strings. Compounds deemed to satisfy the learning objective could be persisted for further examination in the drug discovery process, and the training of the generator could be terminated.
		
		\begin{table}[]
			\centering
			\caption{Experiments hardware specifications}
			\label{tab:hdw_specs}
			\begin{tabular}{|c|c|c|c|c|}
				\hline
				\multicolumn{1}{|c|}{\textbf{Model}}                               & \multicolumn{1}{c|}{\textbf{Number of cores}} & \multicolumn{1}{c|}{\textbf{\begin{tabular}[c]{@{}c@{}}RAM (GB)\end{tabular}}} & \multicolumn{1}{c|}{\textbf{\begin{tabular}[c]{@{}c@{}} GPUs\end{tabular}}} \\ \hline
				\begin{tabular}[c]{@{}l@{}}Intel Xeon \\ CPU E5-2687W\end{tabular} & 48                                     & 128                                                                             & \begin{tabular}[c]{@{}l@{}}1 GeForce\\  GTX 1080\end{tabular}                                                 \\ \hline
				\begin{tabular}[c]{@{}l@{}}Intel Xeon \\ CPU E5-2687W\end{tabular} & 24                                     & 128                                                                             & \begin{tabular}[c]{@{}l@{}}4 GeForce \\ GTX 1080Ti\end{tabular}                                                    \\ \hline
			\end{tabular}
		\end{table}

		\begin{table*}[]
			\centering
			\caption{Performance of ML models used as evaluation functions in the experiments. The reported values are averages of a 5-fold CV in each case. The standard deviation values are shown in parenthesis. The corresponding experiment (s) that used each ML model is/are specified in parenthesis in the first column.}
			\label{tab:ml_metrics}
			\resizebox{\linewidth}{!}{
				\begin{tabular}{|l|l|l|l|l|l|l|}
					\hline
					\multicolumn{1}{|c|}{\multirow{2}{*}{{Model}}} & \multicolumn{4}{c|}{{Binary Classification}}                                                                                                         & \multicolumn{2}{c|}{{Regression}}                                 \\ \cline{2-7} 
					\multicolumn{1}{|c|}{}                                & \multicolumn{1}{c|}{{Precision}} & \multicolumn{1}{c|}{{Recall}} & \multicolumn{1}{c|}{{Accuracy}} & \multicolumn{1}{c|}{{AUC}} & \multicolumn{1}{c|}{{RMSE}} & \multicolumn{1}{c|}{\textbf{$R^2$}} \\ \hline
					RNN-Bin (DRD2)                                        & 0.971 (0.120)                           & 0.970 (0.120)                        & 0.985 (0.011)                          & 0.996 (0.001)                     & -                                  & -                                   \\ \hline
					RNN-Reg (LogP)                                        & -                                       & -                                    & -                                      & -                                 & 0.845 (0.508)                      & 0.708 (0.359)                       \\ \hline
					XGB-Reg (JAK2 Min \& Max)                                        & -                                       & -                                    & -                                      & -                                 & 0.646 (0.037)                      & 0.691 (0.039)                       \\ \hline
				\end{tabular}
			}
		\end{table*}
		
		\subsection{Experiments Setup}\label{sec:exp_setup}
		We performed four experiments to evaluate the approach described in this study. In each experiment, the aim is to ascertain the effectiveness of our proposed approach and how it compares to the case where the reward function is known, as found in the literature. The hardware specifications we used for our experiments are presented in Table~\ref{tab:hdw_specs}. The performance of each of the evaluation functions used in the experiments are shown in Table~\ref{tab:ml_metrics}.
		
		\subsubsection{DRD2 activity}
		This experiment's objective was to train the generator to produce compounds that target the Dopamine Receptor D2 protein. Hence, we retrieved a DRD2 dataset of $351529$ compounds from ExCAPE-DB~\cite{Sun2017}\footnote{\url{https://git.io/JUgpt}}. This dataset contained $8323$ positive compounds (binding to DRD2). We then sampled an equal number of the remaining compounds as the negatives (non-binding to DRD2) to create a balanced dataset. The balanced DRD2 dataset of $16646$ samples was then used to train a two-layer LSTM RNN, similar to the reward function network shown in~\ref{fig:reward_critic_nets}-left but with an additional Sigmoid endpoint, using five-fold cross-validation with the BCE loss function.
		
		The resulting five models of the CV training then served as the evaluation function $E$ of this experiment. This evaluation function is referred to as RNN-Bin in Table~\ref{tab:ml_metrics}. At test time, the average value of the predicted DRD2 activity probability was assigned as the result of the evaluation of a given compound.
		
		Also, to create the set of demonstrations $\mathcal{D}$ of this experiment, we used the SVM classifier of~\cite{Olivecrona2017} to filter the ChEMBL dataset of~\cite{Popova2018} for compounds with a probability of activity greater than $0.8$. This filtering resulted in a dataset of $7732$ compounds to serve as $\mathcal{D}$.
		
		\subsubsection{LogP Optimization}
		In this experiment, we trained a generator to produce compounds biased toward having their octanol-water partition coefficient (logP) less than five and greater than one. LogP is one of the elements of Lipinski's rule of five.
		
		On the evaluation function $E$ of this experiment, we used the LogP dataset of~\cite{Popova2018}, consisting of $14176$ compounds, to train an LSTM RNN, similar to the reward function network shown in~\ref{fig:reward_critic_nets}-left, using five-fold cross-validation with the Mean Square Error (MSE) loss function. Similar to the DRD2 experiment, the five models serve as the evaluation function of $E$. The evaluation function of this experiment is labeled RNN-Reg in Table~\ref{tab:ml_metrics}.
		
		We constructed the LogP demonstrations dataset $\mathcal{D}$ by using the LogP-biased generator of~\cite{Popova2018} to produce $10000$ SMILES strings, of which $5019$ were unique valid compounds. The $5019$ compounds then served as the set of demonstrations for this experiment.
		
		\subsubsection{JAK2 Inhibition}
		We also performed two experiments on producing compounds for JAK2 modulation. In the first JAK2 experiment, we trained a generator to produce compounds that maximize the negative logarithm of half-maximal inhibitory concentration (pIC$_{50}$) values for JAK2. In this instance, we used the public JAK2 dataset of~\cite{Popova2018}, consisting of $1911$ compounds, to train an XGBoost model in a five-fold cross-validation method with the MSE loss function. The five resulting models then served as the evaluation function $E$, similar to the LogP and DRD2 experiments. The JAK2 inhibition evaluation function is referred to as XGB-Reg in Table~\ref{tab:ml_metrics}. We used the JAK2-maximization-biased generator of~\cite{Popova2018} to produce a demonstration set of $3608$ unique valid compounds out of $10000$ generated SMILES.
		
		On the other hand, we performed an experiment to bias the generator towards producing compounds that minimize the pIC$_{50}$ values for JAK2. JAK2 minimization is useful for reducing off-target effects. While we maintained the evaluation function of the JAK2 maximization experiment in the JAK2 minimization experiment, we replaced the demonstrations set with $285$ unique valid SMILES, out of $10000$ generated SMILES, produced by the JAK2-minimization-biased generator of~\cite{Popova2018}. Also, we did not train a reward function for JAK2 minimization but rather transferred the reward function learned for JAK2 maximization to this experiment. However, we negated each reward obtained from the JAK2 maximization reward function for the minimization case study. This was done to examine the ability to transfer a learned reward function to a related domain.
		
		\subsubsection{Model Types}
		As discussed in Section~\ref{sec:prior_model_pretraining}, we pretrained a two-layer Stack-RNN model with the ChEMBL dataset of~\cite{Popova2018} for one epoch. The training time was approximately 14 days. This pretrained model served as the initializing model for the following generators:
		\begin{enumerate}
			\item PPO-GRL: This model type follows our proposed approach. It is trained using the GRL objective at the IRL phase and the PPO algorithm at the RL phase.
			
			\item PPO-Eval: This model type follows our proposed approach but without the IRL phase. The RL algorithm used is PPO. Since the problem presented in section~\ref{sec:problem} assumes an evaluation function $E$ to periodically determine the performance of the generator during training, this model enables us to evaluate an instance where $E$ is able to serve as a reward function directly, such as ML models in our experiments. We note that other instances of $E$, such as molecular docking or a robot performing synthesis, may be expensive to serve as the reward function in RL training.
			
			\item REINFORCE: This model type uses the proposed SMILES generation method in~\cite{Popova2018} to train a two-layer stack-RNN generator following the method and reward functions of~\cite{Popova2018}. Thus, no IRL is performed. In the DRD2 experiment, we used the reward function of~\cite{Olivecrona2017}. Also, for the JAK2 and LogP experiments, we used their respective reward functions in~\cite{Popova2018}.
			
			\item REINFORCE-GRL: Likewise, the REINFORCE-GRL model type is trained using the REINFORCE algorithm at the RL phase and the GRL method to learn the reward function. This model facilitates assessment of the significance of the PPO algorithm properties to our proposed DIRL approach.
			
			\item Stack-RNN-TL: This model type is trained using TL. Specifically, after the pretraining stage, the demonstrations dataset is used to fine-tune the prior model to bias the generator towards the desired chemical space. Unlike the previous model types which are either trained using RL/IRL, this generator is trained using supervised learning (cross entropy loss function). Since this model type is a possible candidate for scenarios where the reward function is not specified but exemplar data could be provided, TL serves as a useful baseline to compare the PPO-GRL approach.
		\end{enumerate}
		The training time of each generator in each experiment is shown in Table~\ref{tab:exp_results_no_threshold}.
		
		\subsubsection{Metrics}\label{sec:metrics}
		Apart from the Stack-RNN-TL model, all other models in each experiment were trained for a maximum of $600$ episodes (across all trajectories), with early termination once the threshold of the experiment was reached. The Stack-RNN-TL model was trained for two epochs (due to time constraint) in each experiment. The threshold for each experiment is the average score of the evaluation function on the demonstration set. Also, only the model weights yielding the best score, as evaluated by $E$ during training, were saved for each model type.
		
		Furthermore, we assessed the performance of all trained generators using the metrics provided by~\cite{Guimaraes2017} and the internal diversity metric in~\cite{Benhenda2017}. In each of these metrics, the best value is 1, and the worst value is 0.
		We give a brief introduction of the metrics below:
		\begin{itemize}
			\item Diversity: The diversity metrics measure the relative diversity between the generated compounds and a reference set. Given a compound from the generated set, a value of 1 connotes that the substructures of the compound is diverse from the referenced set whereas a value of 0 indicates that the compound shares several substructures with the compounds in the reference set. In our study, a random sample of the demonstrations dataset (1000 compounds) constitute the external diversity~\cite{Guimaraes2017} reference set. On the other hand, all generated compounds constitute the reference set when calculating internal diversity~\cite{Benhenda2017}. Intuitively, the internal diversity metric indicates whether the compound generator repeats similar substructures. We used ECFP8 (Extended Connectivity Fingerprint with diameter 8 represented using 2048 bits) vector of each compound for calculating internal and external diversities.
			\item Solubility: Measures the likeliness for a compound to mix with water. This is also referred to as the water-octanol partition coefficient.
			\item Naturalness: Measures how similar a generated compound is to the structure space of Natural Products (NPs). NPs are small molecules that are synthesized by living organisms and are viewed as good starting points for drug development~\cite{Sorokina2019}.
			\item Synthesizability: Measures how a compound lends itself to chemical synthesis (0 indicates hard to make and 1 indicates easy to make)~\cite{Ertl2009}.
			\item Druglikeness: Estimates the plausibility of a generated compound being a drug candidate. The synthesizability and solubility of a compound contribute to the compound's druglikeness.
		\end{itemize}
	
	\begin{figure*}[]
		\centering
		\subcaptionbox{\label{fig:results_drd2}}{
			\includegraphics[width=.4\linewidth]{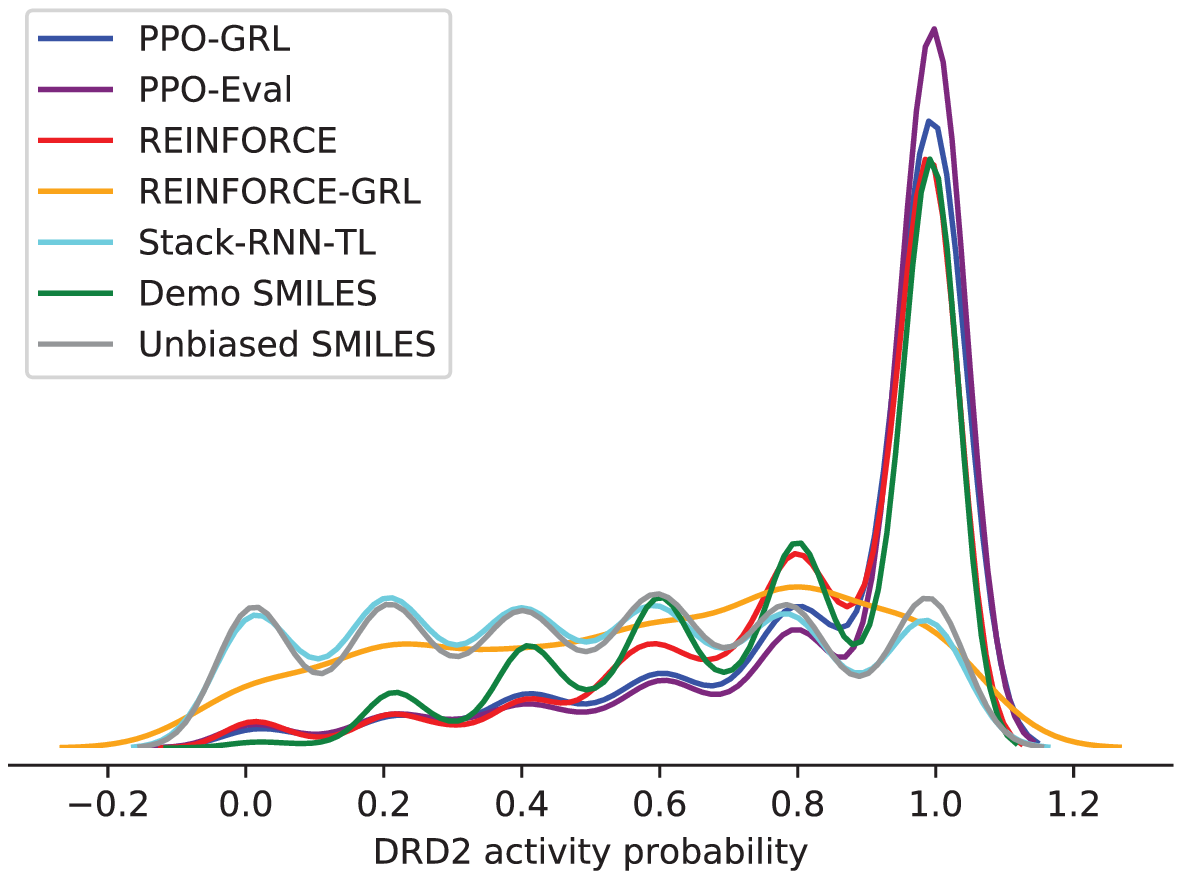}
			\hspace{2em}
			\includegraphics[width=.32\linewidth]{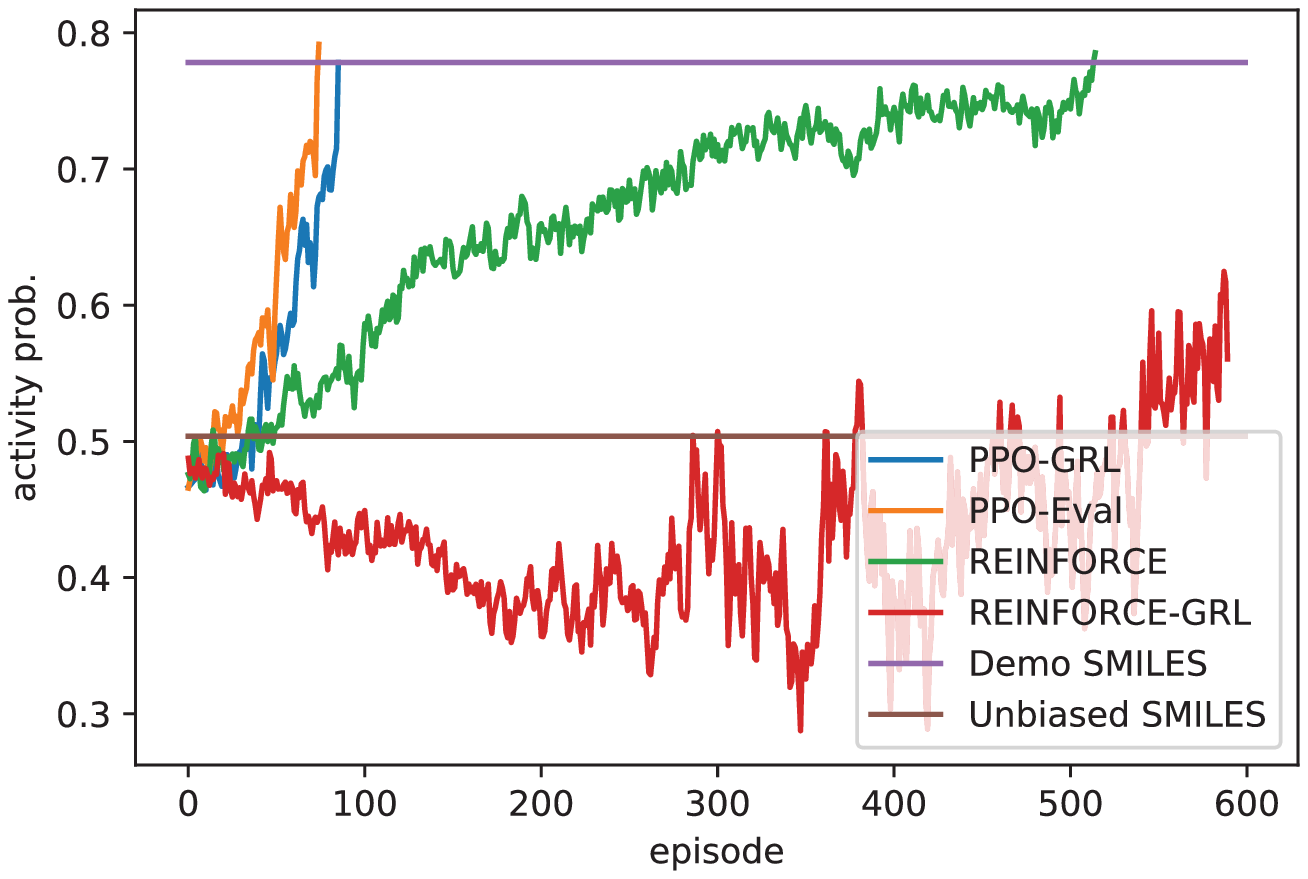}
			\hspace{2em}
			\includegraphics[width=.32\linewidth]{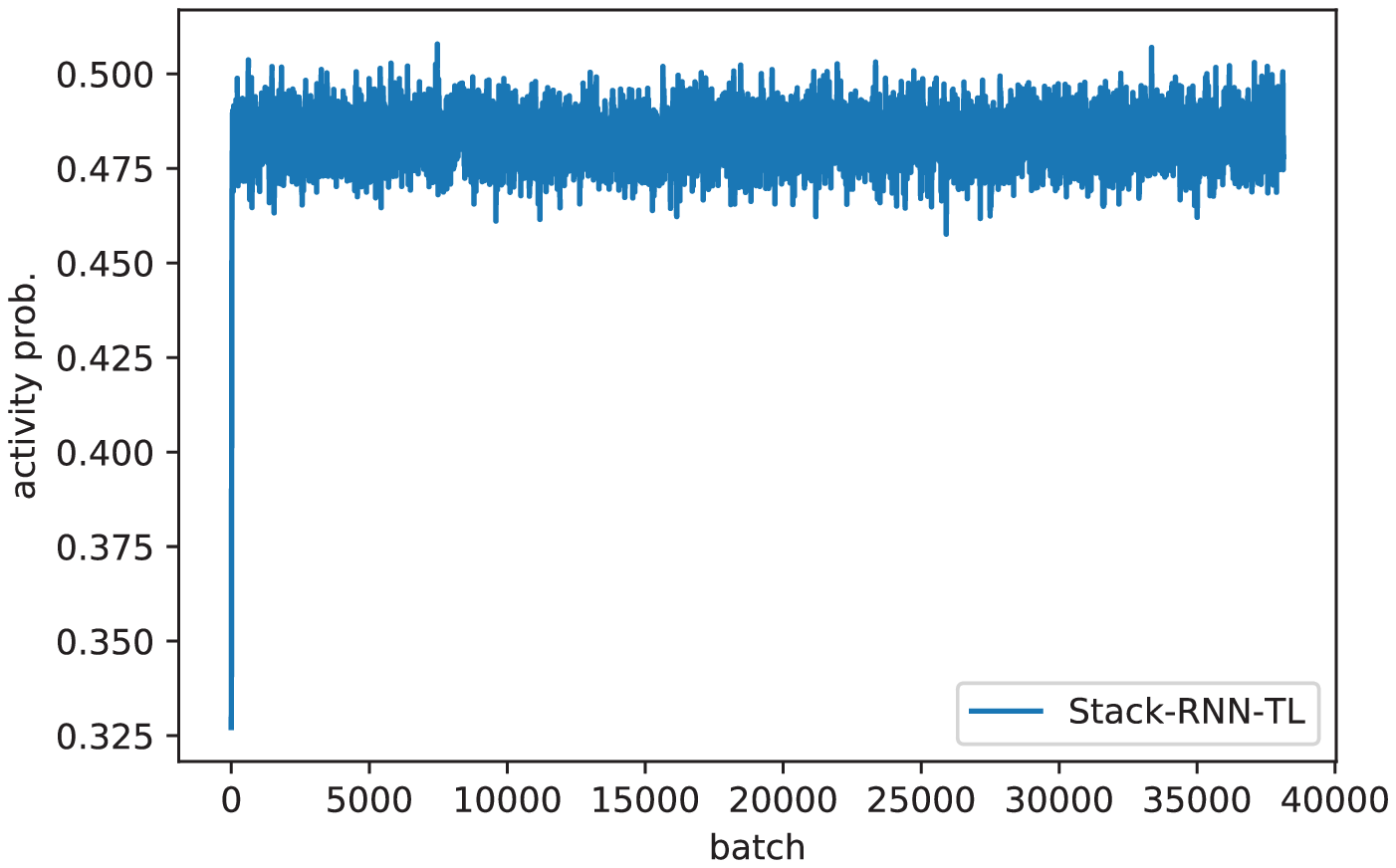}
		}\hspace{1em}%
		\\
		\subcaptionbox{\label{fig:results_logp}}{
			\includegraphics[width=.4\linewidth]{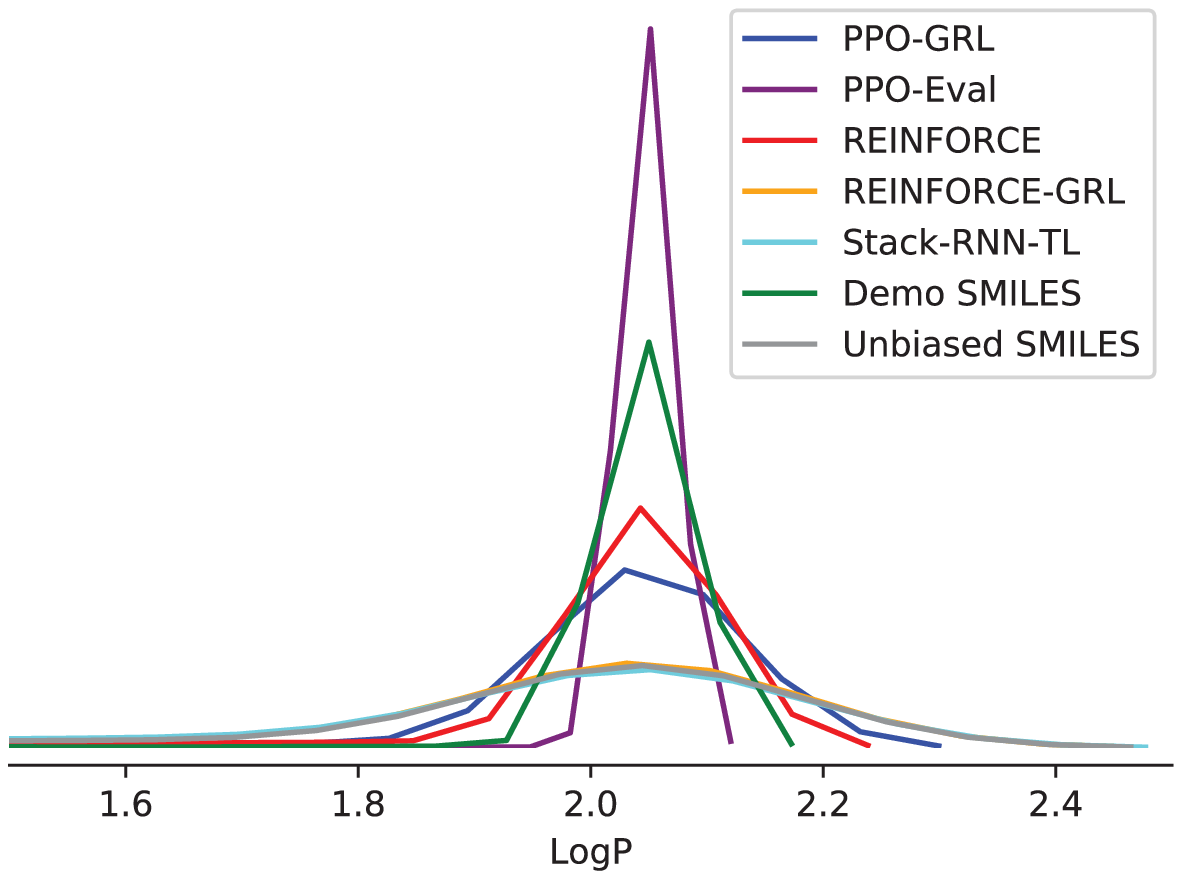}
			\hspace{2em}
			\includegraphics[width=.32\linewidth]{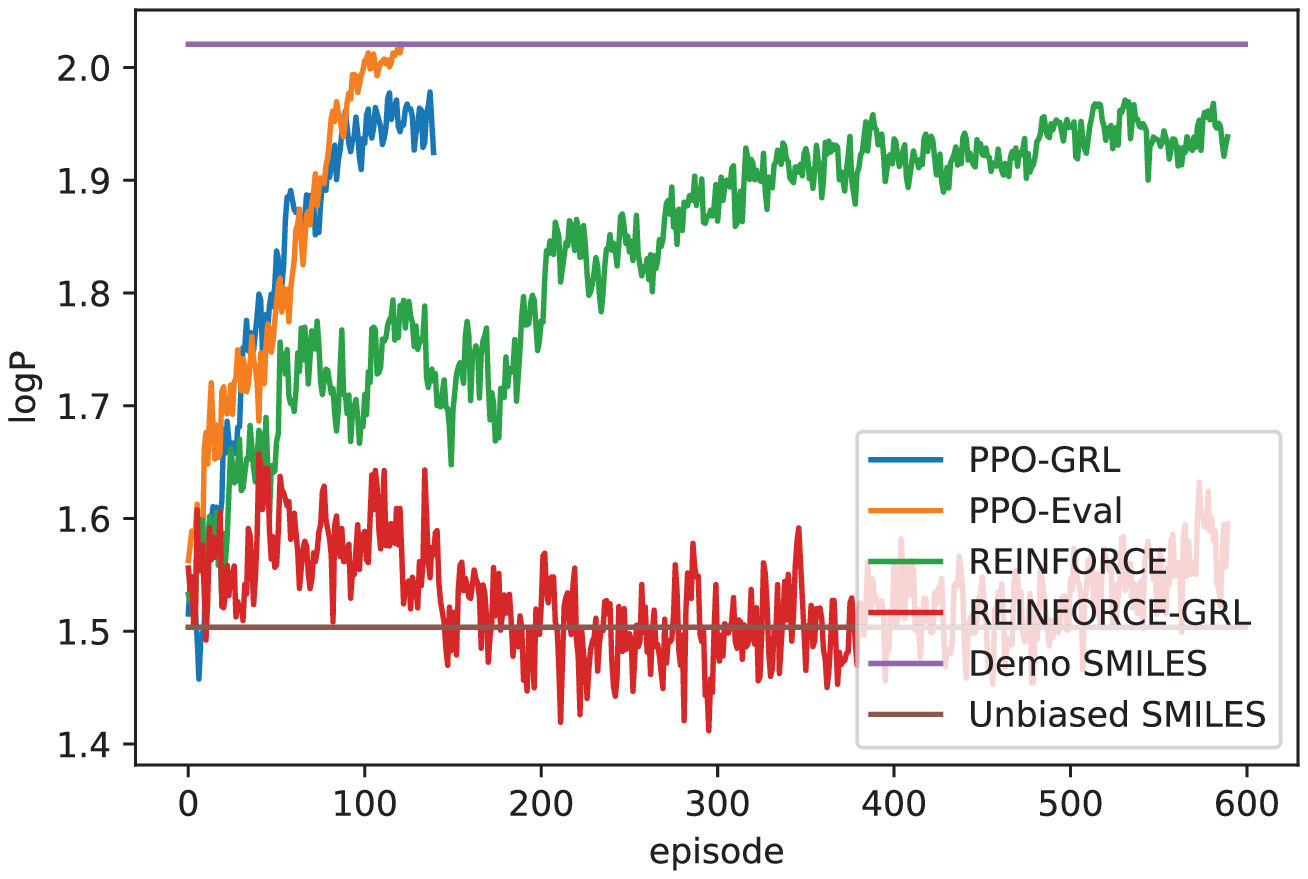}
			\hspace{2em}
			\includegraphics[width=.32\linewidth]{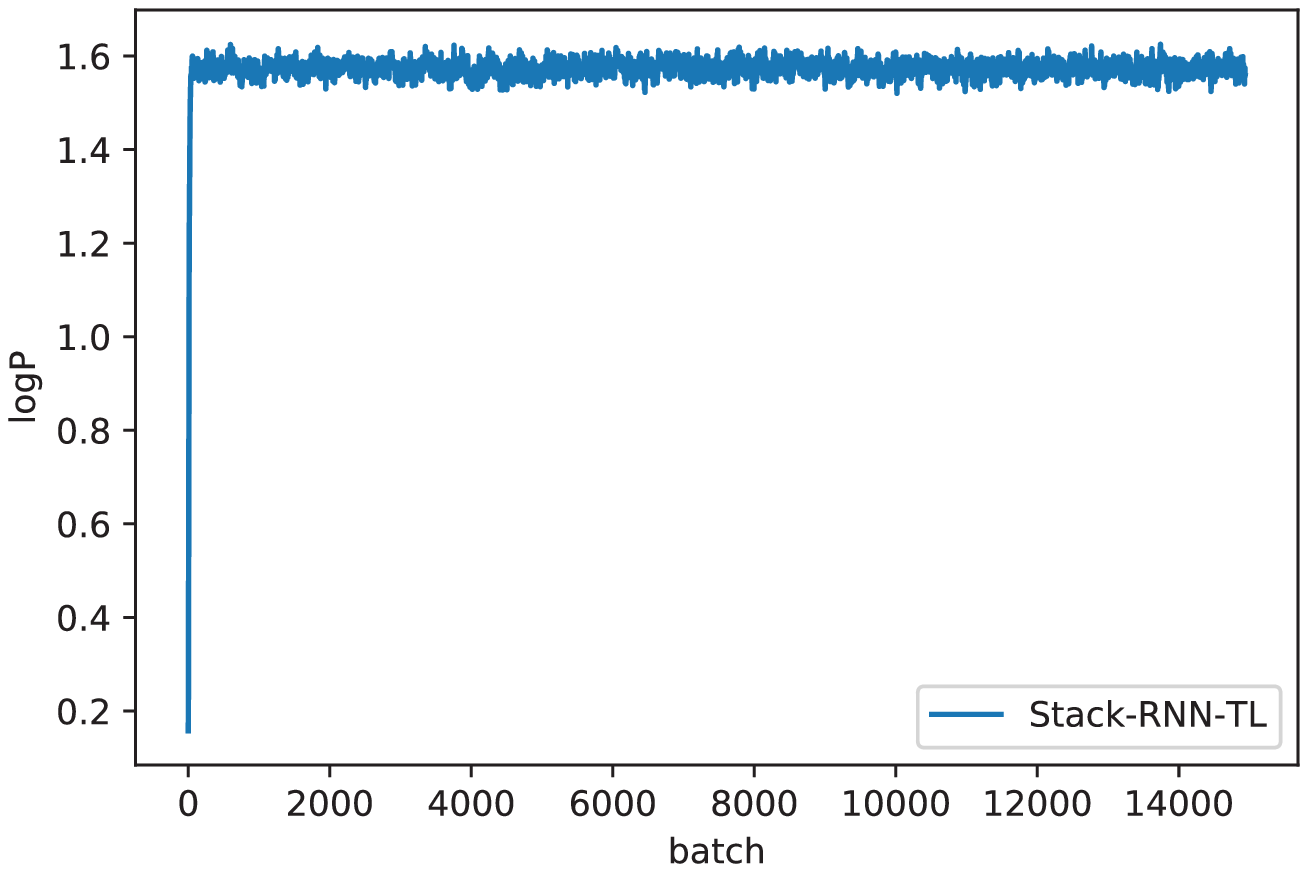}
		}\hspace{1em}%
		\\
		\subcaptionbox{\label{fig:results_jak2_max}}{
			\includegraphics[width=.4\linewidth]{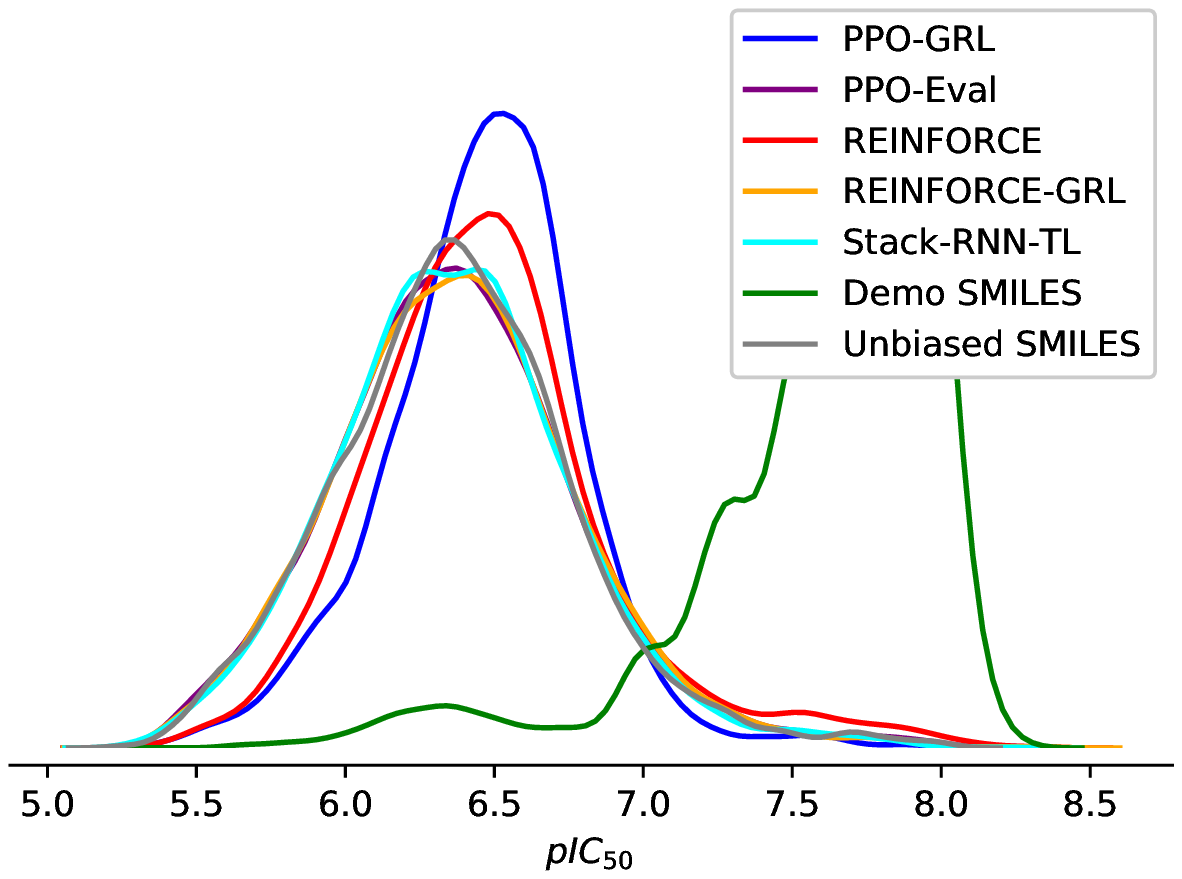}
			\hspace{2em}
			\includegraphics[width=.32\linewidth]{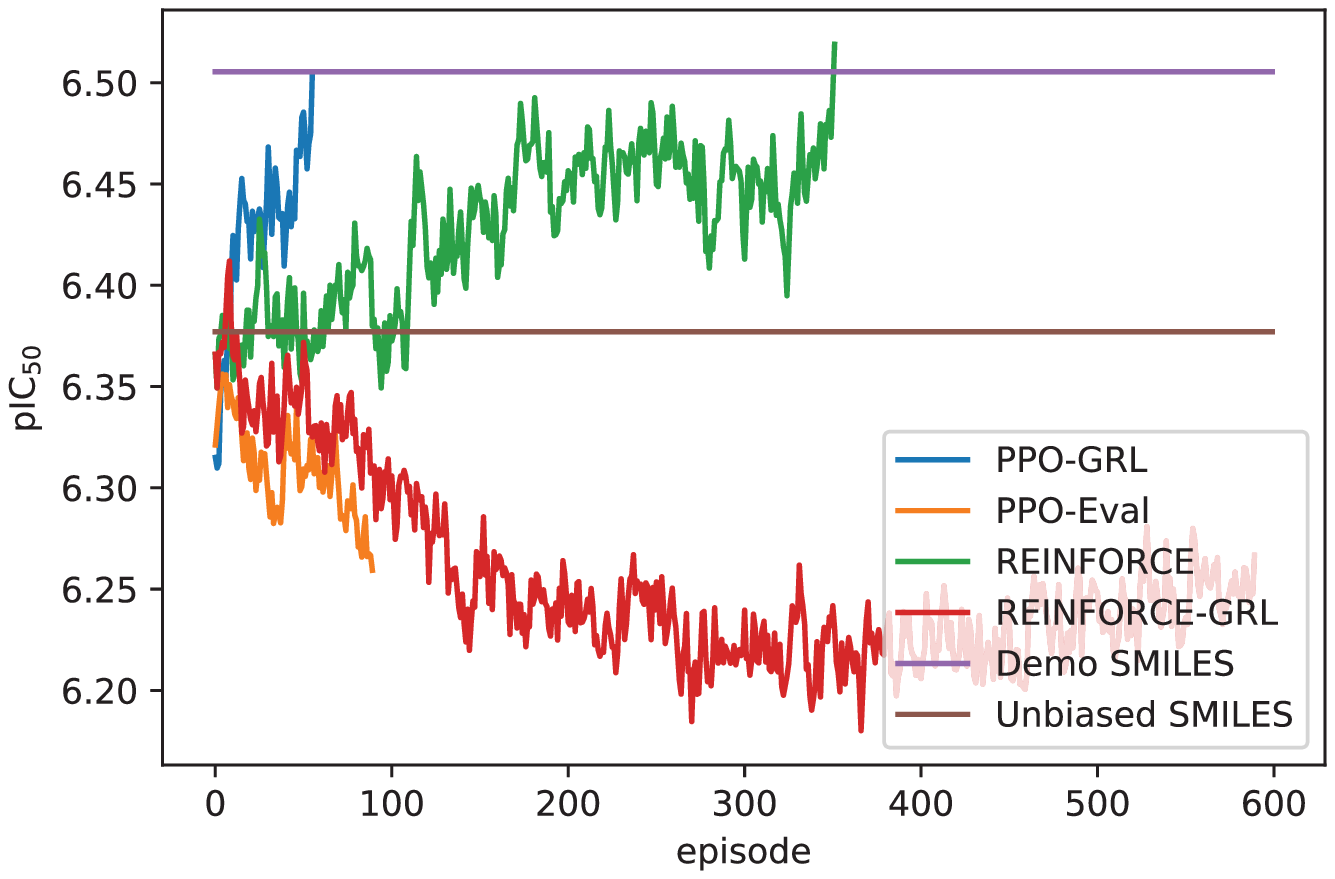}
			\hspace{2em}
			\includegraphics[width=.32\linewidth]{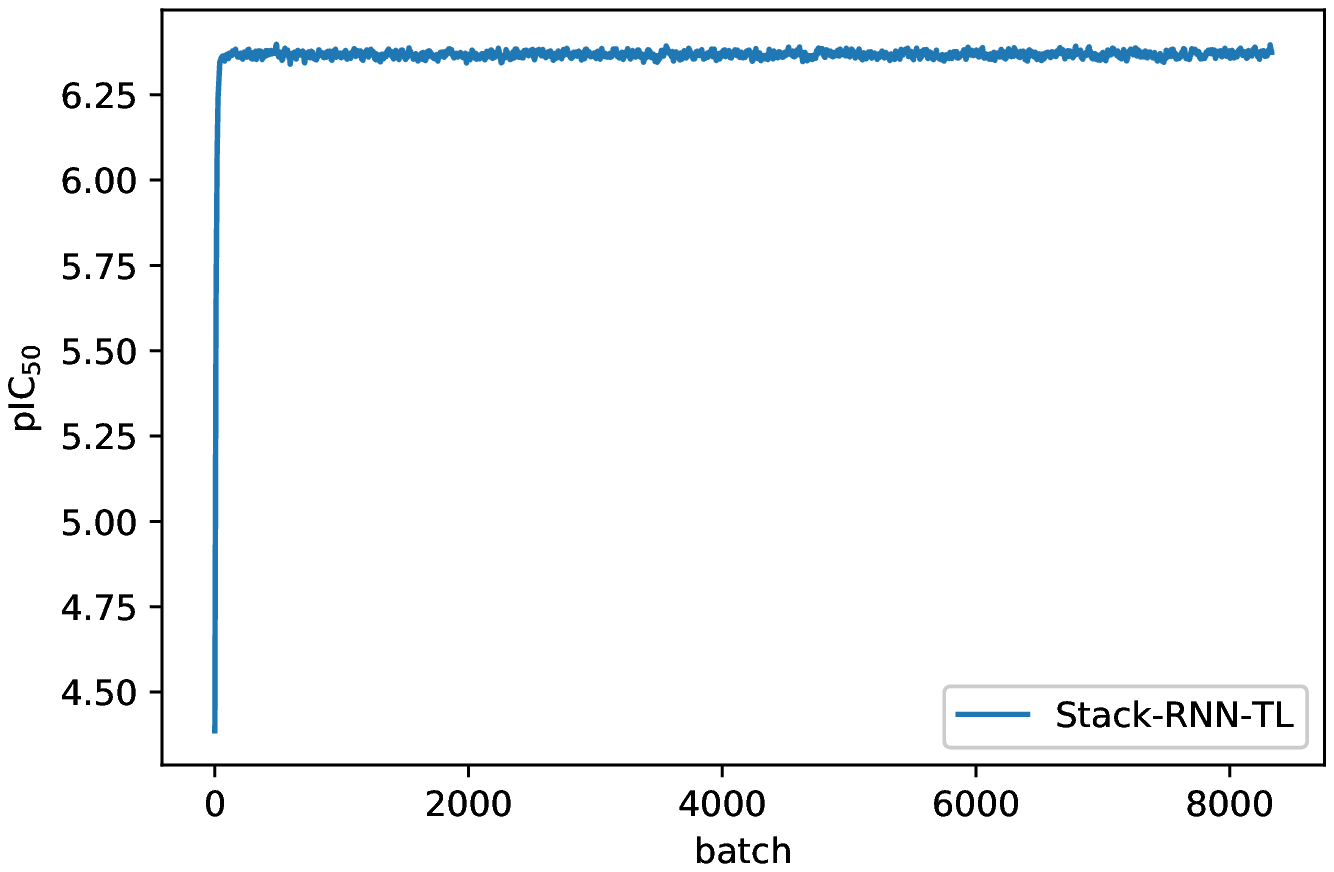}
		}\hspace{1em}%
		\\
		\subcaptionbox{\label{fig:results_jak2_min}}{
			\includegraphics[width=.4\linewidth]{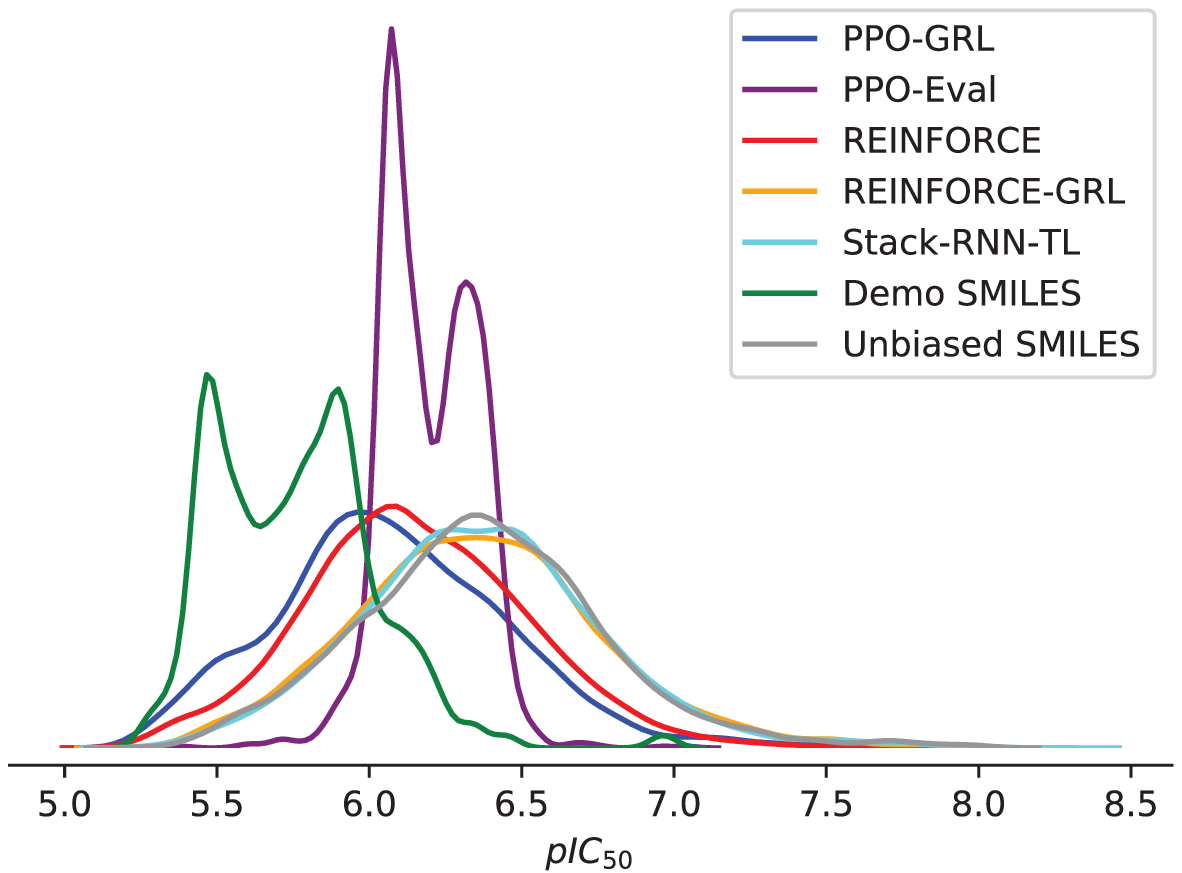}
			\hspace{2em}
			\includegraphics[width=.32\linewidth]{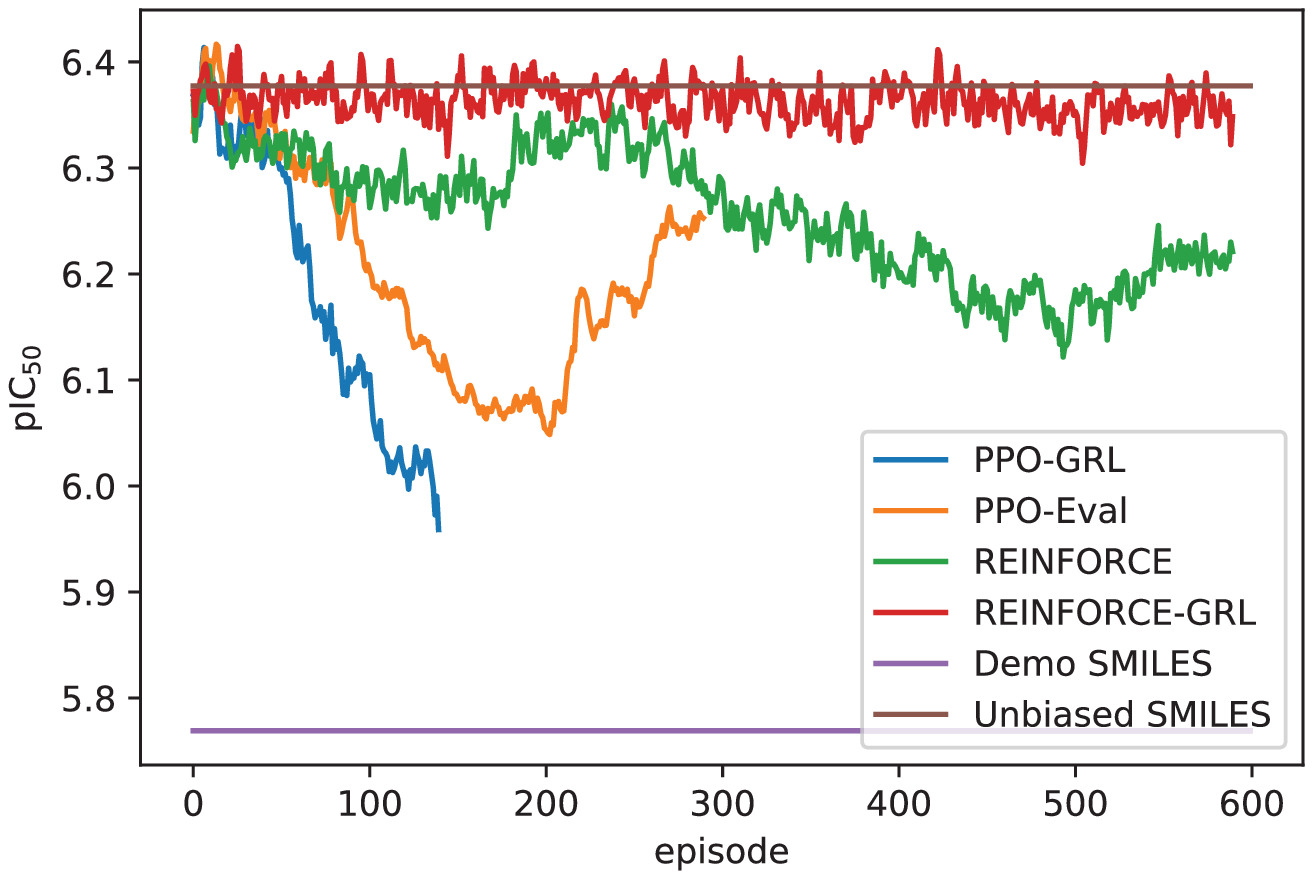}
			\hspace{2em}
			\includegraphics[width=.32\linewidth]{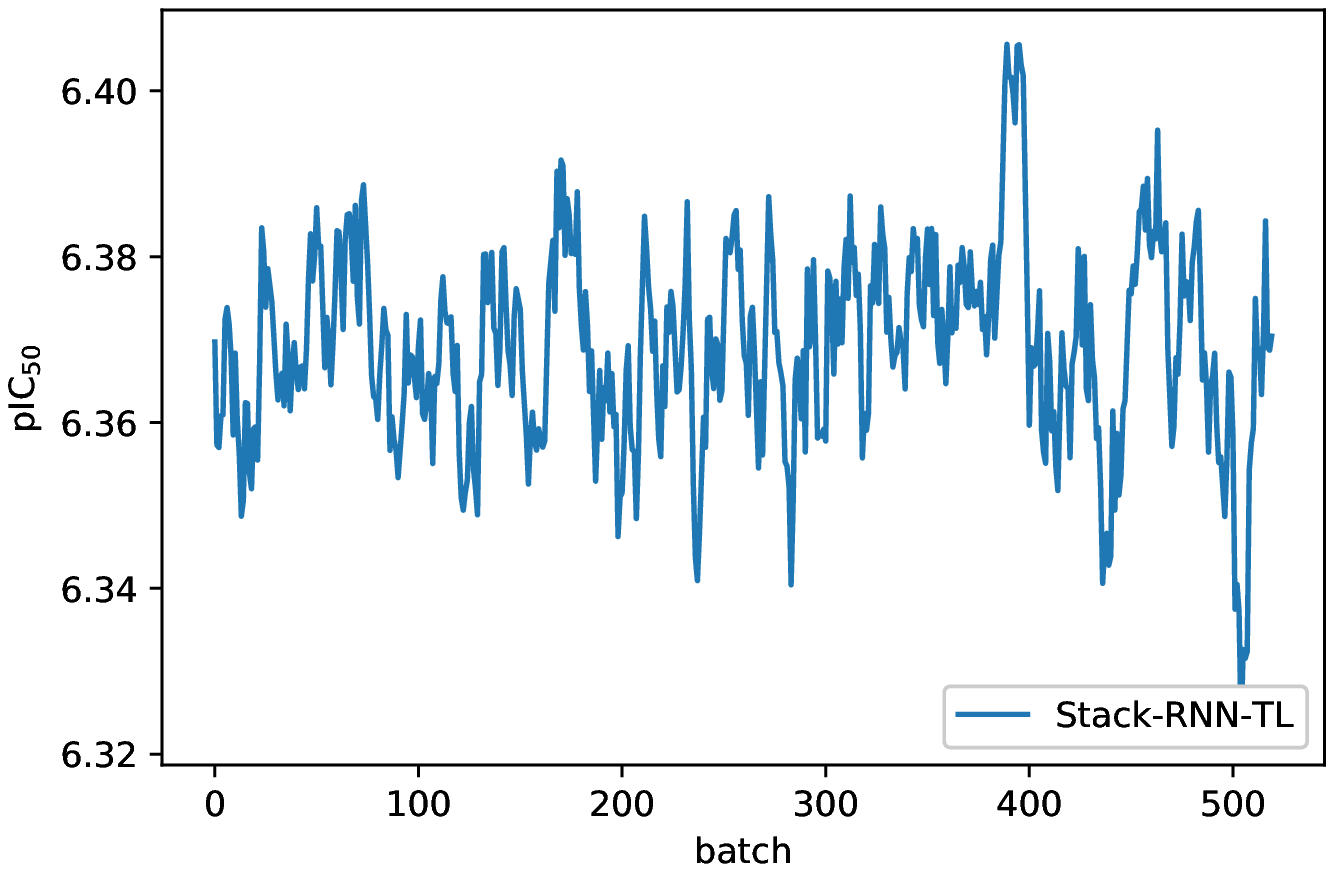}
		}\hspace{1em}%
		
		\caption{\textbf{Column 1} - The distribution plots of the evaluation function's outputs for samples generated from the different model types used in the experiment. \textbf{Column 2} - The convergence plot of the RL and IRL-trained models during training; the y-axis represents the mean value of the experiment's evaluation function output for samples generated at a point during training. The \textit{Demo SMILES} results correspond to the demonstration files of a given experiment. The \textit{Unbiased SMILES} results correspond to samples generated from the pretrained (unbiased or prior) model. \textbf{Column 3} - The convergence plot of the Stack-RNN-TL model during training. (a) The results of DRD2 experiment. (b) The results of LogP optimization experiment. (c) The results of JAK2 maximization. (d) The results of JAK2 minimization experiment.}
		\label{fig:kde_and_conv_plots}
	\end{figure*}

	\begin{table*}[]
		\centering
		\caption{Results of experiments without applying threshold to generated or dataset samples. PPO-GRL follows our proposed approach, REINFORCE follows the work in~\cite{Popova2018}, PPO-Eval, REINFORCE-GRL, and Stack-RNN-TL are baselines in this study.}
		\label{tab:exp_results_no_threshold}
		\resizebox{\textwidth}{!}{
			\begin{tabular}{|l|l|c|c|c|c|c|c|c|c|}
				\hline
				\multicolumn{1}{|c|}{\textbf{Objective}} & \multicolumn{1}{c|}{\textbf{\begin{tabular}[c]{@{}c@{}}Algorithm/\\ Dataset\end{tabular}}} & \textbf{\begin{tabular}[c]{@{}c@{}}Num. of  unique\\ canonical SMILES\end{tabular}} & \textbf{\begin{tabular}[c]{@{}c@{}}Internal \\ Diversity\end{tabular}} & \textbf{\begin{tabular}[c]{@{}c@{}}External \\ Diversity\end{tabular}} & \textbf{Solubility} & \textbf{Naturalness} & \textbf{Synthesizability} & \textbf{Druglikeness} & \textbf{\begin{tabular}[c]{@{}c@{}}Approx. Training\\ Time (minutes)\end{tabular}} \\ \hline
				\multirow{7}{*}{DRD2}                    & Demonstrations                                                                             & 7732                                                                                & 0.897                                                                  & -                                                                      & 0.778               & 0.536                & 0.638                     & 0.603        & -         \\ \cline{2-10} 
				& Unbiased                                                                                   & 2048                                                                                & 0.917                                                                  & 0.381                                                                  & 0.661               & 0.556                & 0.640                     & 0.602       & -          \\ \cline{2-10} 
				& PPO-GRL                                                                                    & 3264                                                                                & 0.887                                                                  & 0.189                                                                  & 0.887               & 0.674                & 0.579                     & 0.271       & 41          \\ \cline{2-10} 
				& PPO-Eval                                                                                   & 4538                                                                                & 0.878                                                                  & 0.183                                                                  & 0.885               & 0.722                & 0.521                     & 0.236    & 33             \\ \cline{2-10} 
				& REINFORCE                                                                                  & 7393                                                                                & 0.904                                                                  & 0.325                                                                  & 0.802               & 0.703                & 0.718                     & 0.443    & 42             \\ \cline{2-10} 
				& REINFORCE-GRL                                                                              & 541                                                                                 & 0.924                                                                  & 0.513                                                                  & 0.846               & 0.794                & 0.293                     & 0.304     &390            \\ \cline{2-10}
				& Stack-RNN-TL         & 6927                                                                                 & 0.918                                                                  & 0.391                                                                  & 0.655               & 0.551                & 0.647                     & 0.611     &3064            \\ \hline
				\multirow{7}{*}{LogP}                    & Demonstrations                                                                             & 5019                                                                                & 0.880                                                                   & -                                                                      & 0.900                 & 0.553                & 0.803                     & 0.512     & -            \\ \cline{2-10} 
				& Unbiased                                                                                   & 2051                                                                                & 0.917                                                                  & 0.372                                                                  & 0.658               & 0.553                & 0.640                      & 0.603    & -             \\ \cline{2-10} 
				& PPO-GRL                                                                                    & 4604                                                                                & 0.903                                                                  & 0.159                                                                  & 0.836               & 0.580                 & 0.770                      & 0.494      & 60           \\ \cline{2-10} 
				& PPO-Eval                                                                                   & 4975                                                                                & 0.733                                                                  & 0.060                                                                   & 0.982               & 0.731                & 0.449                     & 0.075      & 50           \\ \cline{2-10} 
				& REINFORCE                                                                                  & 7704                                                                                & 0.897                                                                  & 0.101                                                                  & 0.774               & 0.580                 & 0.827                     & 0.618       & 56          \\ \cline{2-10} 
				& REINFORCE-GRL                                                                              & 7225                                                                                & 0.915                                                                  & 0.347                                                                  & 0.663               & 0.551                & 0.689                     & 0.627    & 62             \\ \cline{2-10}
				& Stack-RNN-TL                                                                              & 6927                                                                                 & 0.918                                                                  & 0.391                                                                  & 0.655               & 0.551                & 0.647                     & 0.611     &1480            \\ \hline
				\multirow{7}{*}{JAK2 Max}                & Demonstrations                                                                             & 3608                                                                                & 0.805                                                                  & -                                                                      & 0.083               & 0.485                & 0.560                      & 0.483    & -             \\ \cline{2-10} 
				& Unbiased                                                                                   & 2050                                                                                & 0.917                                                                  & 0.379                                                                  & 0.654               & 0.554                & 0.641                     & 0.604        & -         \\ \cline{2-10} 
				& PPO-GRL                                                                                    & 5911                                                                                & 0.928                                                                  & 0.614                                                                  & 0.587               & 0.664                & 0.468                     & 0.581         & 22        \\ \cline{2-10} 
				& PPO-Eval                                                                                   & 6937                                                                                & 0.917                                                                  & 0.386                                                                  & 0.657               & 0.554                & 0.644                     & 0.604     & 38            \\ \cline{2-10} 
				& REINFORCE                                                                                  & 6768                                                                                & 0.916                                                                  & 0.351                                                                  & 0.608               & 0.608                & 0.529                     & 0.627     & 30            \\ \cline{2-10} 
				& REINFORCE-GRL                                                                              & 7039                                                                                & 0.917                                                                  & 0.381                                                                  & 0.658               & 0.555                & 0.644                     & 0.607     & 133            \\ \cline{2-10}
				& Stack-RNN-TL                                                                              & 6927                                                                                 & 0.918                                                                  & 0.391                                                                  & 0.655               & 0.551                & 0.647                     & 0.611     &608            \\ \hline
				\multirow{7}{*}{JAK2 Min}                & Demonstrations                                                                             & 285                                                                                 & 0.828                                                                  & -                                                                      & 0.534               & 0.548                & 0.895                     & 0.604      & -           \\ \cline{2-10} 
				& Unbiased                                                                                   & 2050                                                                                & 0.917                                                                  & 0.379                                                                  & 0.654               & 0.554                & 0.641                     & 0.604       & -          \\ \cline{2-10} 
				& PPO-GRL                                                                                    & 3446                                                                                & 0.907                                                                  & 0.244                                                                  & 0.488               & 0.506                & 0.776                     & 0.663       & 34          \\ \cline{2-10} 
				& PPO-Eval                                                                                   & 1533                                                                                & 0.703                                                                  & 0.008                                                                  & 0.997               & 0.756                & 0.414                     & 0.049      & 148           \\ \cline{2-10} 
				& REINFORCE                                                                                  & 7694                                                                                & 0.908                                                                  & 0.234                                                                  & 0.655               & 0.591                & 0.799                     & 0.649     & 40             \\ \cline{2-10} 
				& REINFORCE-GRL                                                                              & 6953                                                                                & 0.917                                                                  & 0.376                                                                  & 0.662               & 0.547                & 0.657                     & 0.613     & 47            \\ \cline{2-10}
				& 
				Stack-RNN-TL         & 6927                                                                                 & 0.918                                                                  & 0.391                                                                  & 0.655               & 0.551                & 0.647                     & 0.611     &83            \\ \hline
			\end{tabular}
		}
	\end{table*}
	
	\begin{table*}[]
		\centering
		\caption{Results of experiments with applied optimization threshold on generated or dataset samples. PPO-GRL follows our proposed approach, REINFORCE follows the work in~\cite{Popova2018}, PPO-Eval, REINFORCE-GRL, Stack-RNN-TL are baselines in this study.}
		\label{tab:results_with_threshold}
		\resizebox{\linewidth}{!}{
			\begin{tabular}{|l|l|c|l|c|c|c|c|c|c|}
				\hline
				\multicolumn{1}{|c|}{\textbf{Objective}} & \multicolumn{1}{c|}{\textbf{\begin{tabular}[c]{@{}c@{}}Algorithm/\\ Dataset\end{tabular}}} & \textbf{\begin{tabular}[c]{@{}c@{}}Num. of  unique\\ canonical SMILES\end{tabular}} & \multicolumn{1}{c|}{\textbf{\begin{tabular}[c]{@{}c@{}}Proportion \\ in threshold\end{tabular}}} & \textbf{\begin{tabular}[c]{@{}c@{}}Internal \\ Diversity\end{tabular}} & \textbf{\begin{tabular}[c]{@{}c@{}}External \\ Diversity\end{tabular}} & \textbf{Solubility} & \textbf{Naturalness} & \textbf{Synthesizability} & \textbf{Druglikeness} \\ \hline
				\multirow{7}{*}{DRD2}                    & Demonstrations                                                                             & 4941                                                                                & 0.635                                                                                            & 0.896                                                                  & -                                                                      & 0.793               & 0.524                & 0.642                     & 0.588                 \\ \cline{2-10} 
				& Unbiased                                                                                   & 528                                                                                 & 0.266                                                                                            & 0.919                                                                  & 0.412                                                                  & 0.691               & 0.586                & 0.588                     & 0.609                 \\ \cline{2-10} 
				& PPO-GRL                                                                                    & 2406                                                                                & 0.737                                                                                            & 0.864                                                                  & 0.104                                                                  & 0.939               & 0.709                & 0.508                     & 0.181                 \\ \cline{2-10} 
				& PPO-Eval                                                                                   & 3465                                                                                & 0.764                                                                                            & 0.849                                                                  & 0.092                                                                  & 0.947               & 0.756                & 0.458                     & 0.144                 \\ \cline{2-10} 
				& REINFORCE                                                                                  & 5005                                                                                & 0.677                                                                                            & 0.888                                                                  & 0.222                                                                  & 0.837               & 0.715                & 0.706                     & 0.392                 \\ \cline{2-10} 
				& REINFORCE-GRL                                                                              & 201                                                                                 & 0.372                                                                                            & 0.907                                                                  & 0.223                                                                  & 0.929               & 0.853                & 0.211                     & 0.194                 \\ \cline{2-10}
				& Stack-RNN-TL                                                                              & 1715                                                                                 & 0.248                                                                  & 0.920                                                                  & 0.447               & 0.696                & 0.586                     & 0.585     &0.608            \\ \hline
				\multirow{7}{*}{LogP}                    & Demonstrations                                                                             & 5007                                                                                & 1.000                                                                                              & 0.880                                                                   & -                                                                      & 0.898               & 0.553                & 0.803                     & 0.512                 \\ \cline{2-10} 
				& Unbiased                                                                                   & 1903                                                                                & 0.927                                                                                            & 0.915                                                                  & 0.346                                                                  & 0.687               & 0.546                & 0.652                     & 0.605                 \\ \cline{2-10} 
				& PPO-GRL                                                                                    & 4548                                                                                & 0.988                                                                                            & 0.903                                                                  & 0.149                                                                  & 0.842               & 0.579                & 0.772                     & 0.493                 \\ \cline{2-10} 
				& PPO-Eval                                                                                   & 4969                                                                                & 0.999                                                                                            & 0.733                                                                  & 0.061                                                                  & 0.983               & 0.731                & 0.449                     & 0.075                 \\ \cline{2-10} 
				& REINFORCE                                                                                  & 7638                                                                                & 0.991                                                                                            & 0.897                                                                  & 0.098                                                                  & 0.777               & 0.579                & 0.829                     & 0.618                 \\ \cline{2-10} 
				& REINFORCE-GRL                                                                              & 6796                                                                                & 0.941                                                                                            & 0.914                                                                  & 0.317                                                                  & 0.686               & 0.546                & 0.698                     & 0.630                  \\ \cline{2-10}
				& Stack-RNN-TL                                                                              & 6479                                                                                 & 0.935                                                                  & 0.916                                                                  & 0.364               & 0.678                & 0.545                     & 0.655     &0.614            \\ \hline
				\multirow{7}{*}{JAK2 Max}                & Demonstrations                                                                             & 3449                                                                                & 0.958                                                                                            & 0.794                                                                  & -                                                                      & 0.073               & 0.482                & 0.560                      & 0.479                 \\ \cline{2-10} 
				& Unbiased                                                                                   & 717                                                                                 & 0.354                                                                                            & 0.916                                                                  & 0.360                                                                   & 0.639               & 0.572                & 0.565                     & 0.572                 \\ \cline{2-10} 
				& PPO-GRL                                                                                    & 2768                                                                                & 0.468                                                                                            & 0.925                                                                  & 0.543                                                                  & 0.578               & 0.700                  & 0.376                     & 0.533                 \\ \cline{2-10} 
				& PPO-Eval                                                                                   & 2427                                                                                & 0.350                                                                                             & 0.916                                                                  & 0.369                                                                  & 0.640                & 0.564                & 0.565                     & 0.576                 \\ \cline{2-10} 
				& REINFORCE                                                                                  & 2942                                                                                & 0.434                                                                                            & 0.912                                                                  & 0.281                                                                  & 0.586               & 0.609                & 0.462                     & 0.605                 \\ \cline{2-10} 
				& REINFORCE-GRL                                                                              & 2508                                                                                & 0.356                                                                                            & 0.916                                                                  & 0.355                                                                  & 0.644               & 0.561                & 0.568                     & 0.577                 \\ \cline{2-10}
				& Stack-RNN-TL                                                                              & 2376                                                                                 & 0.343                                                                  & 0.917                                                                  & 0.370               & 0.650                & 0.557                     & 0.578     &0.580             \\ \hline
				\multirow{7}{*}{JAK2 Min}                & Demonstrations                                                                             & 145                                                                                 & 0.518                                                                                            & 0.785                                                                  & -                                                                      & 0.539               & 0.506                & 0.908                     & 0.617                 \\ \cline{2-10} 
				& Unbiased                                                                                   & 123                                                                                 & 0.062                                                                                            & 0.887                                                                  & 0.075                                                                  & 0.680                & 0.422                & 0.827                     & 0.655                 \\ \cline{2-10} 
				& PPO-GRL                                                                                    & 693                                                                                 & 0.201                                                                                            & 0.88                                                                   & 0.061                                                                  & 0.534               & 0.453                & 0.852                     & 0.673                 \\ \cline{2-10} 
				& PPO-Eval                                                                                   & 10                                                                                  & 0.007                                                                                            & 0.618                                                                  & 0.000                                                                      & 1.000                   & 0.728                & 0.387                     & 0.050                  \\ \cline{2-10} 
				& REINFORCE                                                                                  & 1078                                                                                & 0.140                                                                                             & 0.891                                                                  & 0.086                                                                  & 0.681               & 0.500                  & 0.887                     & 0.677                 \\ \cline{2-10} 
				& REINFORCE-GRL                                                                              & 490                                                                                 & 0.070                                                                                             & 0.897                                                                  & 0.142                                                                  & 0.682               & 0.432                & 0.831                     & 0.671                 \\ \cline{2-10}
				& Stack-RNN-TL                                                                              & 426                                                                                 & 0.0.061                                                                  & 0.897                                                                  & 0.130               & 0.683                & 0.430                     & 0.832     &0.662            \\ \hline
			\end{tabular}
		}
	\end{table*}

	\begin{figure}
		\centering
		\includegraphics[width=.7\linewidth]{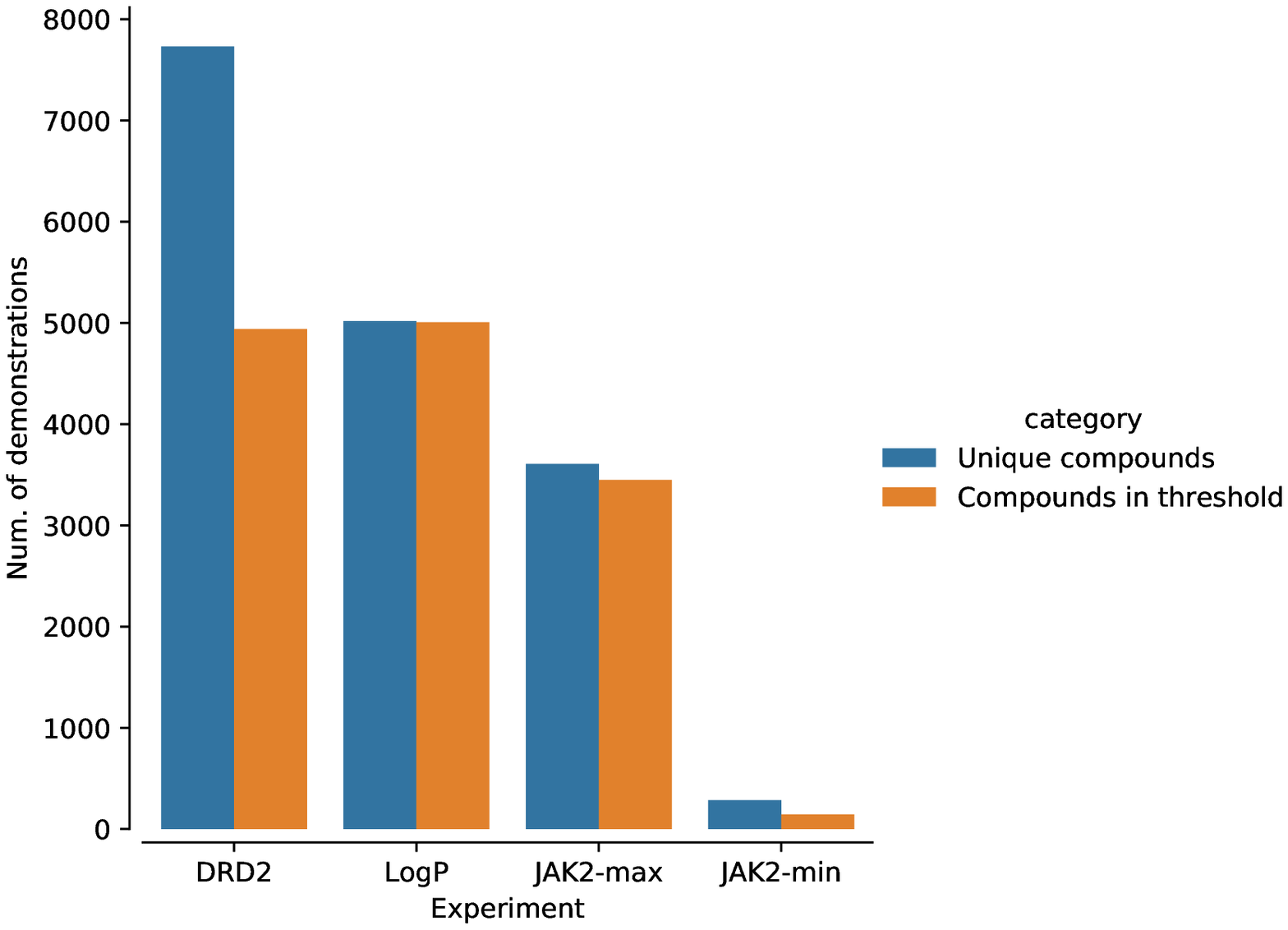}
		\caption{The size of the demonstrations dataset used in each experiment of this study.}
		\label{fig:demosizes}
	\end{figure}
	
	\section{Discussion}\label{sec:discussion}
	We generated $10000$ samples for each trained generator and used the molecular metrics in Section~\ref{sec:metrics} to assess the generator's performance. However, during training, we generated $200$ samples and maintained an exponential average to monitor performance. Figure~\ref{fig:kde_and_conv_plots} shows the density plots of each model's valid SMILES string and the convergence progression of each model toward the threshold of an experiment, beginning from the score of samples generated by the pretrained model.
	
	Also, Table~\ref{tab:exp_results_no_threshold} presents the results of each metric for the valid SMILES samples of each model. Likewise, Table~\ref{tab:results_with_threshold} presents the results for a set of compounds filtered from the valid SMILES samples of each model by applying the threshold of each experiment. In the case of the logP results in Table~\ref{tab:results_with_threshold}, we selected compounds whose values are less than 5 and greater than 1. The added \textit{proportion in threshold} column in Table~\ref{tab:results_with_threshold} presents the quota of a generator's valid SMILES that were within the threshold of the experiment; the maximum value is 1 and minimum is 0.
	
	Firstly, we observed from the results that the PPO-GRL model focused on generating compounds that either satisfy the demonstration dataset threshold or are toward the threshold after a few episodes during training in each of the experiments. This early convergence of the PPO-GRL generator (in terms of number of episodes) connotes that an appropriate reward function has been retrieved from the demonstration dataset to facilitate the biasing optimization using RL. Also, while the evaluation function provided an appropriate signal for training the PPO-Eval in some cases, the diversity metrics of the PPO-Eval model were typically less than the PPO-GRL model. We also realized in training that generating focused compounds toward a threshold was accompanied by an increase in the number of invalid SMILES strings. This increase in invalid compounds explains the drop in the number of valid SMILES of PPO-GRL in Table~\ref{tab:exp_results_no_threshold}. Although the REINFORCE model mostly generated a high number of valid SMILES samples, it was less sample efficient. We reckon that the difference in sample efficiency between PPO-GRL and REINFORCE-GRL is as a result of the variance in estimating~$\Psi$ (see Equation~\ref{eq:policy_gradient}). Thus, a stable estimation of $\Psi$ is significant to the sample-based objective in Equation~\ref{eq:grl_objective} since a better background distribution could be estimated. The performance of the REINFORCE-GRL, as shown in Figure~\ref{fig:kde_and_conv_plots}, reifies this high-variance challenge. Also, the Stack-RNN-TL model recorded the same scores for all metrics across the experiments, as seen in Table~\ref{tab:exp_results_no_threshold}. This performance of the Stack-RNN-TL model connotes that no focused compounds could be generated in each experiment after two epochs of training but mostly took longer time to train as compared to the other models. We also note that the Stack-RNN-TL model produced a higher number of valid canonical SMILES than the unbiased or prior generator after the fine-tuning step.
	
	Concerning the DRD2 experiment, although the PPO-GRL model's number of valid SMILES strings was more than the number of unique compounds in the unbiased dataset, it performed approximately one-third lesser than the PPO-Eval model. As shown in Figure~\ref{fig:results_drd2}, this is observed in the mean predicted DRD2 activity of the PPO-Eval model, which reaches the threshold in fewer episodes than the PPO-GRL and REINFORCE models. The PPO-GRL model produced a higher proportion of its valid compounds in the activity threshold than the REINFORCE model, and the generated samples seem to share a relatively lesser number of substructures with the demonstrations set (external diversity) than the PPO-Eval approach as reported in Table~\ref{tab:results_with_threshold}. Unsurprisingly, due to the variance problem mentioned earlier, the REINFORCE-GRL performed poorly and generated the fewest number of valid compounds with more than half of its produced SMILES being less than the DRD2 activity threshold.
	
	Regarding the logP optimization experiment, most of the compounds sampled from all the generators had logP values that fell within the range of $1$ and $5$. However, while the REINFORCE-GRL and Stack-RNN-TL models recorded an average logP value of approximately $1.6$, the PPO-GRL, PPO-Eval, and REINFORCE models recorded higher average logP values closer to the demonstration dataset's average logP value, as shown in Figure~\ref{fig:results_logp}. Considering that the PPO-GRL model was trained without the reward function used to train the PPO-Eval and REINFORCE generators, our proposed approach was effective at recovering a proper reward function for biasing the generator. Interestingly, the samples of the PPO-GRL generator recorded better diversity scores and were deemed more druglike than the PPO-Eval generator's samples, as shown in Table~\ref{tab:exp_results_no_threshold}.
	
	
	On the JAK2 maximization experiment, the PPO-Eval method could not reach the threshold. The PPO-GRL generator reached the threshold in less than $100$ episodes but with fewer valid SMILES strings than the REINFORCE generator. As shown in Table~\ref{tab:results_with_threshold}, the proportion of compounds in the JAK2 max threshold of the PPO-GRL generated samples, when compared to the other models in the JAK2 maximization experiment, indicates that the PPO-GRL model seems to focus on the objective early in training than the other generators.
	
	On the other hand, the JAK2 minimization experiment provides an insightful view of the PPO-GRL behavior despite its inability, as well as the other models, to reach the threshold in Figure~\ref{fig:results_jak2_min}. In Figure~\ref{fig:demosizes}, we show the size of the demonstrations dataset in each experiment and the number of compounds that satisfy the experiment's threshold. We think of the proportion of each demonstration dataset satisfying the threshold as vital to the learning objective, and hence, fewer numbers could make learning a useful reward function more challenging. Hence, the size and quality of the demonstration dataset contribute significantly to learning a useful reward function. We suggest this explains the PPO-GRL generator's inability to reach the JAK2 minimization threshold. Therefore, it is no surprise that the REINFORCE-GRL generator's best mean pIC$_{50}$ value was approximately the same as that of the pretrained model's score, as seen in Figure~\ref{fig:results_jak2_min}. We note that the number of unique PPO-GRL generated compounds was almost five times larger than the demonstrations with approximately the same internal variation. It is worth noting that while the REINFORCE approach used the JAK2 minimization reward function of~\cite{Popova2018}, the PPO-GRL method used the negated rewards of the learned JAK2 maximization reward function. This ability to transfer learned reward functions could be a useful technique in drug discovery campaigns.
	
	In a nutshell, the preceding results and analysis show that our proposed framework offers a rational approach to train compound generators and learn a reward function, transferable to related domains, in situations where specifying the reward function for RL training is challenging or not possible. While the PPO-Eval model can perform well in some instances, the evaluation function may not be readily available or expensive to serve in the training loop as a reward function in some real-world scenarios such as a robot synthesis or molecular docking evaluation function.

	\section{Conclusion}\label{sec:conclusion}
	This study reviewed the importance of chemical libraries to the drug discovery process and discussed some notable existing proposals in the literature for evolving chemical structures to satisfy specific objectives. We pointed out that while RL methods could facilitate this process, specifying the reward function in certain drug discovery studies could be challenging, and its absence renders even the most potent RL technique inapplicable. Our study has proposed a reward function learning and a structural evolution model development framework based on the entropy maximization IRL method. The experiments conducted have shown the promise such a direction offers in the face of the large-scale chemical data repositories that have become available in recent times. We conclude that further studies into improving the proposed method could provide a powerful technique to aid drug discovery.
	
	\subsection*{\textbf{Further Studies}}
	An area for further research could be techniques for reducing the number of invalid SMILES while the generator focuses on the training objective. An approach that could be investigated is a training method that actively teaches the reward function to distinguish between valid and invalid SMILES strings. Also, a study to provide a principled understanding of developing the demonstrations dataset could be an exciting and useful direction.; Additionally, considering the impact of transformer models in DL research, the Gated Transformer proposed by Parisotto et al.~\cite{Parisotto2019} to extend the capabilities of transformer models to the RL domain offers the opportunity to develop better compound generators. In particular, our work could be extended by replacing the Stack-RNN used to model the generator with a Memory-based Gated Transformer.
	
	\section*{Acknowledgements}
	
	We would like to thank Zhihua Lei, Kwadwo Boafo Debrah, and all reviewers of this study.
	
	\section*{Funding}
	
	This work was partly supported by SipingSoft Co. Ltd.
	\bibliographystyle{unsrt}
	\bibliography{references}
\end{document}